\documentclass{paper}

\usepackage{graphics}
\usepackage{subfigure}
\usepackage{parskip}
\usepackage{color}

\preprintnumber{KA-TP-32-2012\\TTP12-043}

\title{Heavy MSSM Higgs production at the LHC and decays to
  $\boldsymbol{WW,ZZ}$ at higher
orders}

\author{%
  P.\ Gonz\'alez%
  \ref{inst:aachen}\email{gonzalez@physik.rwth-aachen.de}
  \and S. Palmer%
  \ref{inst:kit-itp}%\email{sophy.palmer@particle.uni-karlsruhe.de}
  \and M.\ Wiebusch%
  \ref{inst:kit-ttp}\email{martin.wiebusch@kit.edu}
  \and K.\ Williams%
  \ref{inst:bonn}%\email{k.e.williams@dunelm.org.uk}%
}

\institutes{%
  \institute{inst:aachen}{%
    Institute for Theoretical Particle Physics and Cosmology,\\ 
    RWTH Aachen, D-52056 Aachen, Germany}%
  \institute{inst:kit-itp}{Institute for Theoretical Physics,\\
    Karlsruhe Institute of Technology (KIT), D-76128 Karlsruhe, Germany}%
  \institute{inst:kit-ttp}{Institute for Theoretical Particle Physics,\\
    Karlsruhe Institute of Technology (KIT), D-76128 Karlsruhe, Germany}%
  \institute{inst:bonn}{Department of Physics and Astronomy\\
    University of Bonn, Nußallee 12, D-53115 Bonn, Germany}%
}

\abstract{In this paper we discuss the production of a heavy scalar MSSM Higgs
  boson $H$ and its subsequent decays into pairs of electroweak gauge bosons
  $WW$ and $ZZ$.  We perform a scan over the relevant MSSM parameters, using
  constraints from direct Higgs searches and several low-energy observables.  We
  then compare the possible size of the $pp\to H\to WW,ZZ$ cross sections with
  corresponding Standard Model cross sections. We also include the full MSSM
  vertex corrections to the $H \to WW,ZZ$ decay and combine them with the Higgs
  propagator corrections, paying special attention to the IR-divergent
  contributions. We find that the vertex corrections can be as large as $-30\%$
  in MSSM parameter space regions which are currently probed by Higgs searches
  at the LHC.  Once the sensitivity of these searches reaches two percent of the
  SM signal strength the vertex corrections can be numerically as important as
  the leading order and Higgs self-energy corrections and have to be considered
  when setting limits on MSSM parameters.}

\newcommand{\url}[1]{\texttt{#1}}
\newcommand{\code}[1]{\texttt{#1}}
\newcommand{\Acal}{\mathcal{A}}

\newcommand{\BR}{\text{BR}}

\begin{document}

\input paperdef

\maketitlepage

%%%%%%%%%%%%%%%%%%%%%%%%%%%%%%%%%%%%%%%%%%%%%%%%%%%%%%%%%%%%%%%%%%%%%%%%%%%%%%%%
\section{Introduction}
%%%%%%%%%%%%%%%%%%%%%%%%%%%%%%%%%%%%%%%%%%%%%%%%%%%%%%%%%%%%%%%%%%%%%%%%%%%%%%%%

The recent discovery of a \unit{126}{GeV} resonance decaying into photons and
(off-shell) $Z$ bosons at the LHC \cite{ATLAS:2012gk, CMS:2012gu} opens a new
era in particle physics. The next important task for both theorists and
experimentalists is to determine the exact nature of that resonance. Currently
the measured signals are in statistical agreement with the expectations from a
Standard Model (SM) Higgs boson. However, the experimental sensitivity is not
yet sufficient to rule out an extended Higgs sector, especially if the
(tree-level) couplings of the additional Higgs bosons to electroweak gauge
bosons are suppressed. The discovery of additional scalar resonances would give
us important clues about the exact mechanism of electroweak symmetry breaking.

In the minimal supersymmetric SM (MSSM) the tree-level couplings of the neutral
Higgs bosons to weak gauge bosons are determined by the Higgs mass scale (either
$M_{A}$, the mass of the pseudoscalar Higgs boson, or $M_{H^{\pm}}$, the mass of
the charged Higgs boson) and $\tan\beta$, the ratio of the vacuum expectation
values of the Higgs doublets.  At leading order the masses of the Higgs bosons
are also determined by these two parameters. For $M_{A}\gg M_Z$, the so-called
decoupling limit \cite{Gunion:2002zf,Haber:2002ey}, the heavy scalar Higgs boson
$H$ and the pseudoscalar $A$ are almost degenerate and their (effective)
couplings to $W$ and $Z$ bosons are strongly suppressed. This makes the search
for heavy MSSM Higgs bosons more difficult than the search for a Standard Model
Higgs boson with similar mass. However, it is well-known that the masses and
couplings of MSSM Higgs bosons receive large corrections at higher orders in
perturbation theory
\cite{Ellis:1990nz,Heinemeyer:1998jw,Degrassi:2002fi,Allanach:2004rh}.  Also,
the production rates for Higgs bosons are modified in the MSSM, especially in
the $gg\to H,A$ and $b\bar b\to H,A$ production modes.

In \cite{Bernreuther:2010uw} the production and decays of a pseudoscalar Higgs
into electroweak gauge bosons were discussed in a number of different models,
including the MSSM. In this paper we answer the question of how large the LHC
signal cross sections for $pp\to H\to WW,ZZ$ can become in the MSSM when higher
order corrections to both the production and decay processes are taken into
account. For this purpose we perform a scan over the relevant MSSM parameters,
using experimental constraints from several low-energy observables and direct
Higgs searches at LEP, Tevatron and LHC. We do not assume a specific SUSY
breaking scenario, but scan directly over the soft SUSY breaking parameters at
the electroweak scale. For this scan we make extensive use of the public codes
\code{HiggsBounds 3.8.0} \cite{arXiv:0811.4169,arXiv:1102.1898} and
\code{FeynHiggs 2.7.4} \cite{9812320, 9812472, 0212020, 0611326, 07050746,
  CPC180, Hahn:2010te}. 

The Higgs--gauge-boson couplings are implemented in \code{FeynHiggs} in the
\emph{improved Born-approximation}, i.e.\ taking into account higher order
corrections from Higgs self-energies but no genuine vertex corrections.  The
MSSM vertex corrections for both the $WW$ and $ZZ$ final state were calculated
in \cite{Hollik:2011xd}, although for the $WW$ final state only fermion and
sfermion contributions were considered. For our analysis we performed an
independent calculation of all one-loop vertex corrections and found agreement
with \cite{Hollik:2011xd}. We then extended the analysis of the $H \rightarrow
WW$ case to the complete MSSM corrections, including the IR divergent
contributions and the corresponding real emission graphs. Our scan shows that
the vertex corrections typically lie between $-10\%$ and $-30\%$ in MSSM
parameter space regions where the $H\to WW,ZZ$ channels should still be
observable at the LHC.

The case of off-shell decays of the light MSSM Higgs-boson $h$ was discussed in
\cite{Hollik:2010ji}, where a calculation of the process $h\to W^*W^*,Z^*Z^*\to
4f$ (four fermions) was presented. In this paper we examine the possibility of
calculating the single off-shell processes $H\to WW^*\to Wff'$ and $H\to ZZ^*\to
Wff'$ process in an effective coupling approximation, i.e.\ by re-scaling the
corresponding SM decay rates. Such an approximation can be useful in parameter
scans or fits, where off-shell decays of the heavy MSSM Higgs boson may be of
interest, but a numerical integration of the full four-particle phase space is
not feasible.  We discuss the quality of the approximation and address the issue
of infrared divergences in this approach.

In Section~\ref{sec:details} we introduce our notation and explain the
combination of the vertex corrections with the self-energy corrections
calculated by \code{FeynHiggs}. In Section~\ref{sec:parscan} we give the
details of the parameter scan and discuss the experimental constraints that were
used in it. The numerical results of the scan and the quality of the effective
coupling approximation are discussed in Section~\ref{sec:results}. Our
conclusions are given in Section~\ref{sec:conclusions}.

%%%%%%%%%%%%%%%%%%%%%%%%%%%%%%%%%%%%%%%%%%%%%%%%%%%%%%%%%%%%%%%%%%%%%%%%%%%%%%%%
\section{Details of the calculation}
\label{sec:details}
%%%%%%%%%%%%%%%%%%%%%%%%%%%%%%%%%%%%%%%%%%%%%%%%%%%%%%%%%%%%%%%%%%%%%%%%%%%%%%%%

%%%%%%%%%%%%%%%%%%%%%%%%%%%%%%%%%%%%%%%%%%%%%%%%%%%%%%%%%%%%%%%%%%%%%%%%%%%%%%%%
\subsection{Notation and conventions}
%%%%%%%%%%%%%%%%%%%%%%%%%%%%%%%%%%%%%%%%%%%%%%%%%%%%%%%%%%%%%%%%%%%%%%%%%%%%%%%%

In the MSSM, the Higgs sector contains two scalar doublets, which give five
physical Higgs bosons. At lowest order, the Higgs sector is $\cp$-conserving,
containing two charged Higgs bosons, $H^{\pm}$, two neutral $\cp$-even Higgs
bosons, $h$ and $H$, and the $\cp$-odd Higgs $A$. Two independent parameters
characterise the Higgs sector, normally taken as $M_{A}$ and $\tan \beta$, where
$\tan \beta$ is the ratio of the vacuum expectation values of the Higgs
doublets.  Higher order corrections lead to large corrections to the Higgs
masses and mixing angle $\alpha$, and can induce $\cp$-violation and mixing
between the three neutral Higgs bosons $h$, $H$ and $A$
\cite{Pilaftsis:1998dd,Pilaftsis:1999qt,Pilaftsis:2000au} if complex
SUSY-breaking parameters are allowed.\footnote{In the case of $\cp$-violation,
  it is usual to take $M_{H^{\pm}}$ as input parameter instead of $M_{A}$
  because, in the $\cp$-violating case the pseudoscalar Higgs boson $A$ mixes
  with the $\cp$-even neutral Higgs bosons.}

%%%%%%%%%%%%%%%%%%%%%%%%%%%%%%%%%%%%%%%%%%%%%%%%%%%%%%%%%%%%%%%%%%%%%%%%%%%%%%%%
\subsection{Higgs propagator corrections}
\label{sec:Hprop}
%%%%%%%%%%%%%%%%%%%%%%%%%%%%%%%%%%%%%%%%%%%%%%%%%%%%%%%%%%%%%%%%%%%%%%%%%%%%%%%%

Higgs propagator corrections can be extremely important numerically, especially
in the non-decoupling regions of the SUSY parameter space, and are in addition
needed in order to ensure correct on-shell properties of $S$-matrix elements
involving external Higgs bosons -- i.e.\ unit residue and vanishing mixing
between different Higgs bosons on mass shell.  These corrections can be included
by using finite wave function normalisation factors. In the case where these
factors are applied to a tree level decay amplitude we speak of an
\emph{improved Born approximation}. In the following, quantities computed in
this approximation are denoted with a subscript `imp.B'. An amplitude
$\Acal_{H,\text{imp.B}}$ in the improved Born approximation with an external
Higgs boson $H$ can receive corrections from three tree-level amplitudes
$\Acal_{h,\text{tree}}$, $\Acal_{H,\text{tree}}$ and $\Acal_{A,\text{tree}}$
involving the three neutral Higgs states:
\begin{align}
\label{eq:hprop}
  \Acal_{H,\text{imp.B}}
  =  \matr{Z}_{Hh} \Acal_{h,\text{tree}}
   + \matr{Z}_{HH} \Acal_{H,\text{tree}}
   + \matr{Z}_{HA} \Acal_{A,\text{tree}}
\end{align}
The matrix $\matr{Z}$ has been defined in Ref.~\cite{0611326,07105320} and is
non-unitary.  When no $\cp$-violation is present mixing occurs only between
the $\cp$-even states, but when complex parameters are allowed mixing between
all three neutral states needs to be considered.\footnote{For the $H \rightarrow
VV$ decays considered in this paper, $\Acal_A^\text{tree}$ is of course
zero.}  The program \code{FeynHiggs 2.7.4}~\cite{9812320, 9812472, 0212020,
0611326, 07050746, CPC180, Hahn:2010te} has been used to calculate both the
corrected Higgs boson masses and the wave function normalisation
$\matr{Z}$-factors. \code{FeynHiggs} includes the complete one-loop corrections
as well as the dominant two-loop contributions in the MSSM with real and complex
parameters. 

%% \begin{figure}[!htb]
%% \begin{center}
%%           \resizebox{0.76\hsize}{!}{\includegraphics*{pictures/propINloop}}
%% \end{center}
%%      \caption{Applying the Higgs propagator corrections at both tree and loop
%%        level.}
%% \label{fig:prop}
%% \end{figure}

Since the Higgs propagator corrections are universal, they can in principle be
applied to the loop diagrams as well as the tree-level diagrams. Denoting the
one-loop vertex corrections to the decay amplitudes of the tree-level mass
eigenstates $h$, $H$ and $A$ as $\Delta\Acal_h$, $\Delta\Acal_H$
and $\Delta\Acal_A$, respectively, we define the \emph{improved vertex
corrections} for the physical mass eigenstate as
\begin{equation}
  \Delta\Acal_\text{imp}
  =  \matr{Z}_{Hh} \Delta\Acal_h
   + \matr{Z}_{HH} \Delta\Acal_H
   + \matr{Z}_{HA} \Delta\Acal_A
  \eqsep.
\end{equation}
When computing interferences between the tree-level and one-loop vertex
diagrams, improved versions can be used for neither, the tree-level or both of
the factors.  This provides an easy method of including (potentially large)
higher-order corrections in our calculations.  In the $\cp$-violating case,
applying the propagator corrections at loop level could give rise to interesting
effects as it allows the $\cp$-odd Higgs boson, $A$ (which of course does not
couple to the gauge bosons at tree level), to be taken into account.

In any case, applying the Higgs propagator corrections means that we are mixing
perturbative orders and could potentially miss cancellations found at higher
orders. However, estimations of the uncertainties from unknown higher order
corrections (see \cite{Degrassi:2002fi, Allanach:2004rh, Heinemeyer:2004gx})
indicate that the $\matr{Z}$-factors do indeed give rise to a leading
contribution which is not expected to be numerically compensated by the
remaining 2-loop pieces.  Since the effect of applying the Higgs propagator
corrections at loop level is significant (as we shall show), we choose to follow
this method.

%%%%%%%%%%%%%%%%%%%%%%%%%%%%%%%%%%%%%%%%%%%%%%%%%%%%%%%%%%%%%%%%%%%%%%%%%%%%%%%%
\subsection{Comparing the SM and MSSM}
\label{sec:effc}
%%%%%%%%%%%%%%%%%%%%%%%%%%%%%%%%%%%%%%%%%%%%%%%%%%%%%%%%%%%%%%%%%%%%%%%%%%%%%%%%

In this paper we are interested in MSSM scenarios that lead to relatively large
$pp\to H\to VV$ signals ($VV=WW,ZZ$). Hence, we define the ratios
\begin{align}\label{eq:RVV}
  R_{VV} &= \frac{\sigma_H^\text{MSSM}\cdot\BR(H\to VV)^\text{MSSM}}%
                 {\sigma_H^\text{SM}\cdot\BR(H\to VV)^\text{SM}}
  = \frac{\sigma_H^\text{MSSM}}{\sigma_H^\text{SM}}\cdot
    \frac{\Gamma_H^\text{SM}}{\Gamma_H^\text{MSSM}}\cdot\rho_{VV}
  \eqsep,\nonumber\\
  \rho_{VV} &= \frac{\Gamma(H\to VV)^\text{MSSM}}{\Gamma(H\to VV)^\text{SM}}
  \eqsep.
\end{align}
where $\sigma_H$ denotes the LHC production cross section for the Higgs 
$H$, $\BR(H\to~VV)$ and $\Gamma(H\to VV)$ are the branching ratio and partial
decay width into vector bosons $V$ $(V=W,Z)$ and $\Gamma_H$ is the total decay
width of the Higgs boson $H$.  The superscript `MSSM' indicates that the
corresponding quantity is evaluated in the MSSM, with $H$ being the heavy scalar
MSSM Higgs boson. The superscript `SM' means that the quantity is evaluated in
the SM, with $H$ being a SM Higgs boson with the same mass as the heavy MSSM
Higgs boson. At leading order the ratios $R_{VV}$ and $\rho_{VV}$ are the same
for $V=W$ and $V=Z$, since the ratio between the Standard Model and MSSM
couplings is the same for both $HWW$ and $HZZ$. From the definitions of
Eq.~\eqref{eq:RVV} it is obvious that the $pp\to H\to VV$ cross sections and
$H\to VV$ partial widths within the MSSM can be obtained by scaling the
corresponding SM quantities with $R_{VV}$ or $\rho_{VV}$.

If the Higgs mass $M_H$ is below the $VV$ threshold the Higgs boson $H$ may
still decay into $Vff'$ (with $f$ and $f'$ being light fermions) via an
off-shell vector boson $V^*$. If new-physics contributions to the $Vff'$ vertex
and non-factorisable contributions are neglected, the corresponding ratios of
partial widths or cross sections times branching ratios do not depend on the
fermions $f$ and $f'$. For off-shell decays we therefore define $R_{Vff'}$ and
$\rho_{Vff'}$ as ratios of differential cross sections and partial widths:
\begin{align}
  \rho_{Vff'}(M_{ff'})
  &= \frac{\partial\Gamma(H\to VV^*\to Vff')^\text{MSSM}/\partial M_{ff'}}%
          {\partial\Gamma(H\to VV^*\to Vff')^\text{SM}/\partial M_{ff'}}
  \nonumber\\
  R_{Vff'}(M_{ff'})
  &= \frac{\sigma_H^\text{MSSM}}{\sigma_H^\text{SM}}\cdot
     \frac{\Gamma_H^\text{SM}}{\Gamma_H^\text{MSSM}}\cdot\rho_{Vff'}(M_{ff'})
  \eqsep,
\end{align}
where $M_{ff'}$ denotes the invariant mass of the $ff'$
pair. \emph{Differential} $pp\to H\to Vff'$ cross sections time branching ratios
and \emph{differential} $H\to Vff'$ partial widths within the MSSM may thus be
obtained by scaling the corresponding SM quantities with $R_{Vff'}$ and
$\rho_{Vff'}$. Usually, $\rho_{Vff'}$ is only weakly dependent on $M_{ff'}$.
We may then approximate $\rho_{Vff'}(M_{ff'})$ as a constant,
\begin{equation}\label{eq:effc_approx}
  \rho_{Vff'}(M_{ff'})\approx \rho_{Vff'}(M_H-M_V)\equiv \rho_{Vff'}
  \eqsep,
\end{equation}
and calculate \emph{integrated} MSSM cross sections and partial widths by
scaling corresponding SM quantities with the appropriate factors. This
approximation is what we call the \emph{effective coupling approximation}, since
higher order corrections to the $HVV$ vertex have been absorbed into an
effective coupling constant.

The principle behind this effective coupling approximation is the same as that
used by the Higgs Cross Section Working Group when working in the MSSM.  In
order to include all known higher order corrections (some of which are known
only in the SM, not the MSSM), the Working Group takes SM `building blocks' and
dresses them with the appropriate MSSM coupling factors, as described in
\cite{Dittmaier:2012vm}.

%%%%%%%%%%%%%%%%%%%%%%%%%%%%%%%%%%%%%%%%%%%%%%%%%%%%%%%%%%%%%%%%%%%%%%%%%%%%%%%%
\subsection{Higher-order corrections and form factors}
%%%%%%%%%%%%%%%%%%%%%%%%%%%%%%%%%%%%%%%%%%%%%%%%%%%%%%%%%%%%%%%%%%%%%%%%%%%%%%%%

We can incorporate the corrections to the $HVV$ vertex by calculating an
effective $HVV$ coupling resulting from the loop and counterterm
diagrams. The structure of this coupling for on-shell particles is
\cite{0403297,Kniehl:1991xe,Kniehl:1990mq}
\begin{equation}\label{eq:formfac}
  T^{\mu \nu} (q_{1}, q_{2})
  = A(q_{1},q_{2}) g^{\mu \nu} + B(q_{1},q_{2}) q_{1}^{\mu} q_{2}^{\nu} +
    C(q_{1},q_{2}) \epsilon^{\mu \nu \rho \sigma} q_{1 \rho} q_{2 \sigma}
  \eqsep.
\end{equation}
Here, $q_{1}$ and $q_{2}$ are the momenta of the electroweak gauge bosons, and
$A$, $B$ and $C$ are Lorentz invariant form factors.  For off-shell particles,
the coupling can have a more complicated structure, but if the gauge bosons
decay into massless fermions the only relevant form factors are $A$, $B$ and
$C$. At tree level, only the formfactor $A$ has a non-zero value in both the SM
and the MSSM:
%
% \begin{gather}
%   A^{\sm}_{H_{\sm}WW} = \frac{i\, eM_{W}}{\sin\theta_{W}}  
%   \label{eq:a1SM}\eqsep,\eqsep A^{\sm}_{H_{\sm}ZZ} = \frac{i\, eM_{W}}{\sin\theta_{W}\cos^2\theta_{W}}\\
%   A^{\mssm}_{hWW} = \frac{i\, eM_{W}}{\sin\theta_{W}}
%   \sin(\beta - \alpha)\eqsep,\eqsep 
%   A^{\mssm}_{HWW}  = \frac{i\, eM_{W}}{\sin\theta_{W}} \cos(\beta -
%   \alpha)
%   \eqsep, \nonumber\\
% A^{\mssm}_{hZZ} = \frac{i\, eM_{W}}{\sin\theta_{W}\cos^2\theta_{W}}
%   \sin(\beta - \alpha)\eqsep,\eqsep 
%   A^{\mssm}_{HZZ}  = \frac{i\, eM_{W}}{\sin\theta_{W}\cos^2\theta_{W}} \cos(\beta -
%   \alpha)
%   \eqsep.
% \label{eq:a1MSSM}
% \end{gather}
\begin{gather}
  A^{\sm}_{H_{\sm}WW} = \frac{i\, eM_{W}}{\sin\theta_{W}}  
  \label{eq:a1SM}\eqsep,\eqsep A^{\sm}_{H_{\sm}ZZ} = \frac{i\, eM_{W}}{\sin\theta_{W}\cos^2\theta_{W}}\\
  A^{\mssm}_{hVV} = A^{\sm}_{H_{\sm}VV}
  \sin(\beta - \alpha)\eqsep,\eqsep 
  A^{\mssm}_{HVV}  = A^{\sm}_{H_{\sm}VV} \cos(\beta -
  \alpha)
  \eqsep, VV = WW, ZZ
\label{eq:a1MSSM}
\end{gather}

At lowest order the MSSM formfactor $A$ representing the coupling of the light
$\cp$-even Higgs boson differs from the SM value of $A$ by a factor of
$\sin{\left(\beta - \alpha\right)}$, which tends to~$1$ in the decoupling
regime, i.e.\ for $\MA \gg \MZ$. Higher order diagrams, however, lead to
different contributions to $A$ in the Standard Model and MSSM, and can result in
non-zero values for $B$ and $C$. 

For the calculation of the form factors we employ a mixed renormalisation scheme
where the electroweak sector is renormalised on-shell \cite{Denner}, while the
Higgs sector is renormalised using a hybrid scheme where the Higgs fields are
renormalised in the $\overline{DR}$ scheme and $M_{A}$ is renormalised
on-shell, as described in \cite{0611326}.  We parameterise our results in terms
of $\alpha (M_{Z})$ and calculate the charge renormalisation constant
accordingly -- i.e.
\begin{align}
 \delta Z_{e} \rightarrow \delta Z_{e} - \frac{1}{2} \Delta \alpha
\end{align}
For Higgs bosons inside loops we use the physical masses and the unitary Higgs
mixing matrix calculated by \code{FeynHiggs}, as described in \cite{0611326}.

%%%%%%%%%%%%%%%%%%%%%%%%%%%%%%%%%%%%%%%%%%%%%%%%%%%%%%%%%%%%%%%%%%%%%%%%%%%%%%%%
\subsubsection{Infrared divergences in $H \rightarrow WW$}
\label{subsec:IRdiv}
%%%%%%%%%%%%%%%%%%%%%%%%%%%%%%%%%%%%%%%%%%%%%%%%%%%%%%%%%%%%%%%%%%%%%%%%%%%%%%%%

In the tensorial structure given in Eq.~\eqref{eq:formfac}, it is only the form
factor $A$ that is IR-divergent due to photon exchange. To render transition
probabilities finite one must sum over all energy-degenerate final and initial
states \cite{Lee:1964is,Kinoshita:1962ur}. In practice this means that one has
to include contributions involving real radiation of a photon in order to obtain
infrared-finite observables. In the SM the analytic expression for the on-shell
real correction to the partial width $H_\sm\to W^+W^-$ reads
\begin{align}\label{eq:realrad}
  \Gamma_{3}(H_{SM}\rightarrow WW)
  &=\Gamma_0 \frac{\alpha}{\pi}\cdot\left[
      - F_{1}(\beta_0)\log\frac{\varepsilon}{\beta_0}
      + 2\left(2-\log[2]\right)
      + \frac{1+\beta_0^2}{\beta_0}\left[
          \spence\frac{2\beta_0}{\beta_0-1}\right.\right.
  \\
  &-\left.\left.\spence\frac{2\beta_0}{\beta_0+1}
   +\spence\frac{1+\beta_0}{2}
   -\spence\frac{1-\beta_0}{2}
   +\frac{1}{2}\log[1-\beta_0^2]\log\frac{1+\beta_0}{1-\beta_0}\right]\right]
  \nonumber\\
  &+\frac{M_{H_{\sm}}^3\alpha}{16 \sin\theta_{W}^2 M_W^2}
    \frac{\alpha}{\pi}\left(\left
        (\beta_0^4 + \beta_0^2 - \beta_0^6 - 1\right)
        \log\frac{1+\beta_0}{1-\beta_0}
      + 2\beta_0^5 + 2\beta_0 - \frac{4}{3}\beta_0^3\right),\nonumber
\end{align}
with
\begin{equation}\label{eq:beta0}
  \beta_0 =\sqrt{1-4M_W^2/M_{H_\sm}^2}
\end{equation}
and $M_{H_\sm}$ the mass of the SM Higgs boson.  The function
$F_1(\beta_0)$ is given by
\begin{equation}
  F_{1}(\beta_0)=\frac{1+\beta_0^2}{\beta_0}\log\frac{1+\beta_0}{1-\beta_0}-2
  \eqsep,
\end{equation}
and $\Gamma_0$ denotes the lowest order SM decay width, given by
\begin{equation}
    \Gamma_0
  = \frac{\alpha M_{H_\sm}}{16 \sin\theta_{W}^2}
    \frac{\beta_0}{1-\beta_0^2}\left(3\beta_0^4-2\beta_0^2+3\right)
  \eqsep.
\end{equation}
When $\Gamma_3(H\rightarrow WW)$ is added to the correction to the partial
width from loop diagrams, $\Gamma_\virt(H\rightarrow WW)$, the virtual
corrections cancel the IR divergences in the real corrections order-by-order, as
required by the Bloch-Nordsieck theorem \cite{52PhysRev54}.

To improve the accuracy of our results in the MSSM we would like to apply the
Higgs propagator $\matr{Z}$-factors to the vertex diagrams as well as the tree
diagrams, as described in Section~\ref{sec:Hprop}. However, as discussed
previously, we are then mixing different orders of perturbation theory and the
cancellation of IR divergences is no longer guaranteed.
\begin{figure}[!htb]
\begin{center}
     \subfigure[IR-divergent loop corrections to the $HWW$ vertex.]{
         \label{fig:IR_loop}
         \resizebox{0.75\hsize}{!}{\includegraphics*{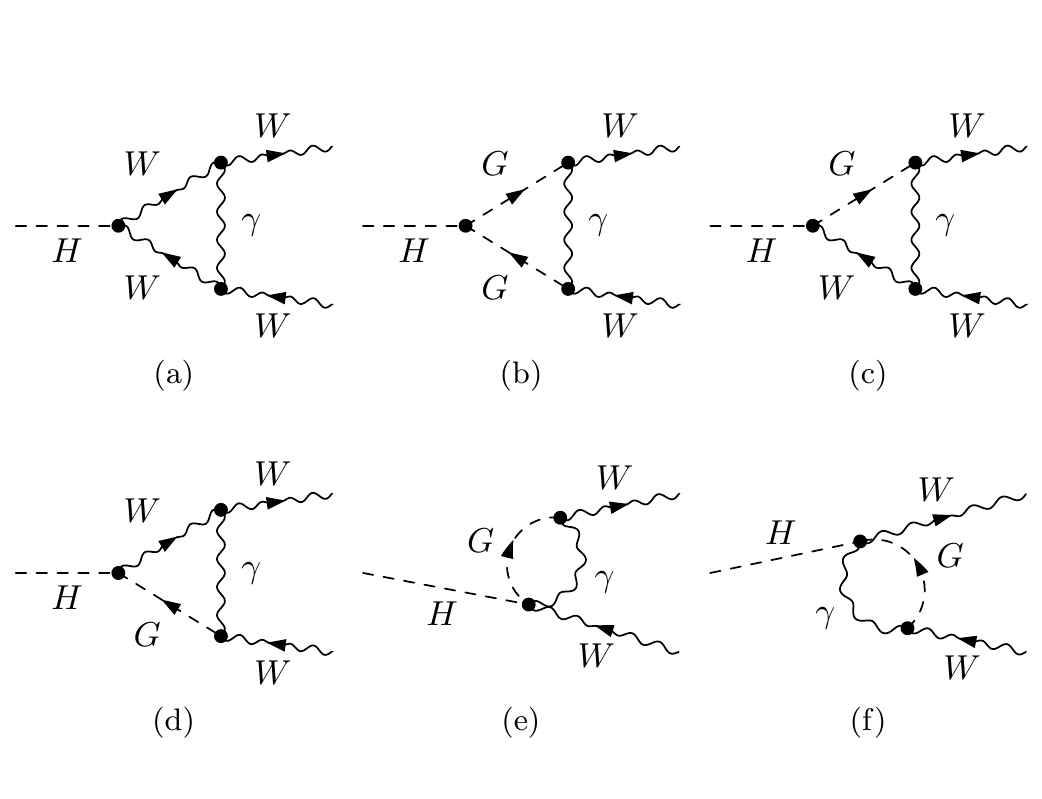}} }
     \\ \subfigure[IR-divergent contributions to the renormalisation constants
       contained within the $HWW$ counterterm.]{
         \label{fig:IR_ct}
         \resizebox{0.45\hsize}{!}{\includegraphics*{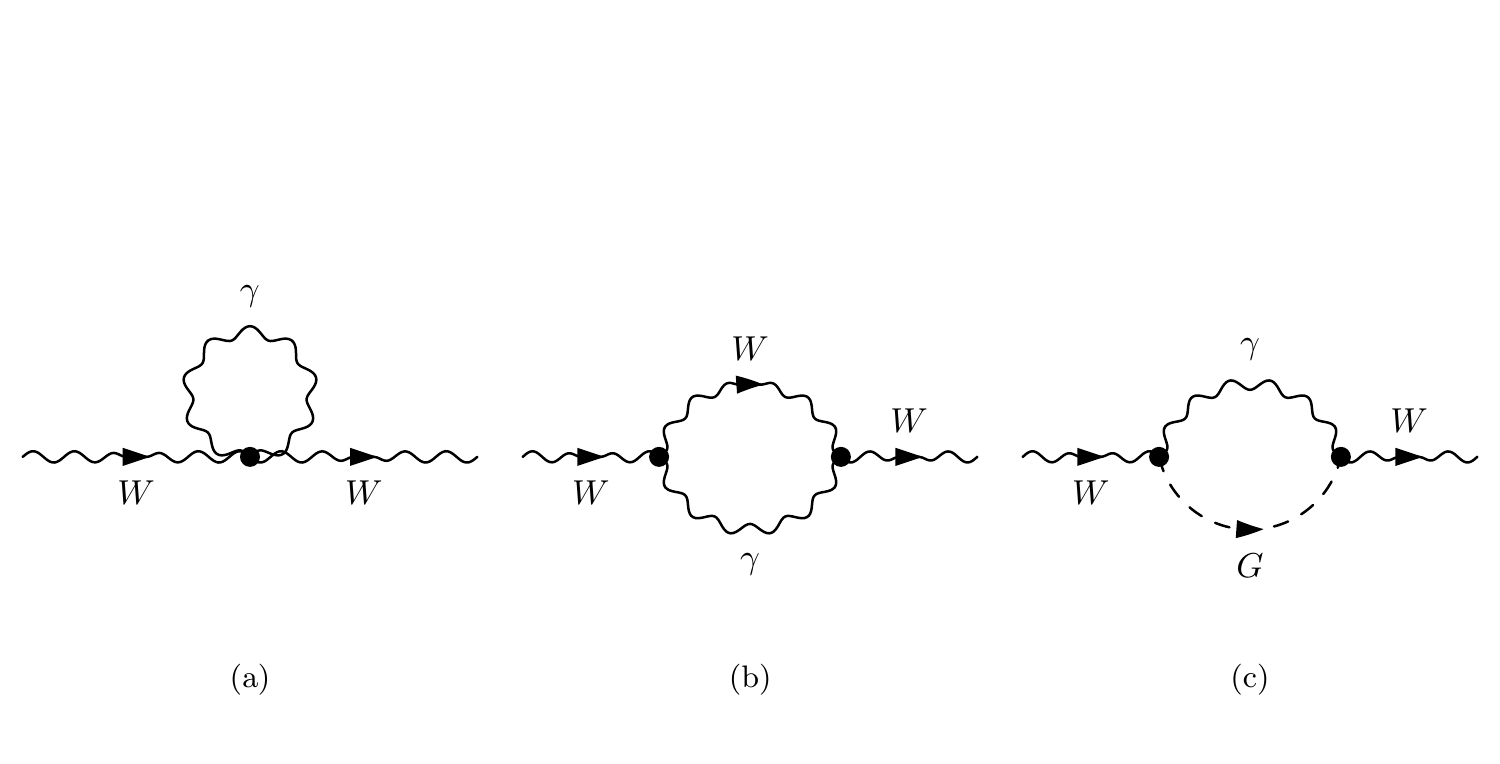}}
         }
     \hspace{.3cm}
     \subfigure[Real corrections to the $HWW$ vertex.]{
         \label{fig:IR_real}
         \resizebox{0.45\hsize}{!}{\includegraphics*{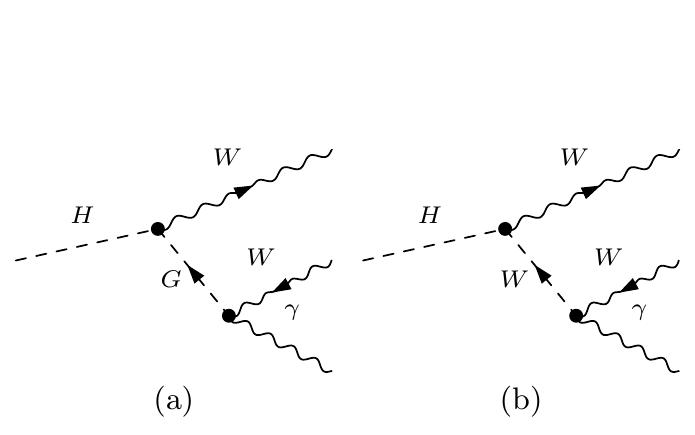}}
         } 
\end{center}
     \caption{IR divergent diagrams contributing to the corrections to the $H
       \rightarrow WW$ partial width.}
     \label{fig:diag_IR}
\end{figure} 
Once the Higgs propagator corrections have been applied according to
Eq.~\eqref{eq:hprop}, the leading order SM and MSSM couplings are related
through the following equation:
\begin{align} 
     A_\impB
  &= \left(\sin(\beta-\alpha) \matr{Z}_{Hh}
     + \cos(\beta-\alpha) \matr{Z}_{HH} \right) A^\sm_\tree
  \nonumber\\
  &= F_\mssm A^\sm_\tree
  \eqsep. 
\end{align}
Here and for the rest of this section, the superscript `SM' indicates that the
corresponding quantity is evaluated in the SM with the SM Higgs boson mass set
to the mass of the heavy MSSM Higgs boson $H$. Symbols without superscripts
refer to the MSSM unless stated otherwise.  When propagator-type corrections are
applied to the loop and real radiation diagrams, the IR divergent NLO diagrams
(shown in Figure~\ref{fig:diag_IR}) are also modified. Counterterm diagrams
(involving renormalisation constants with IR divergent contributions, as shown
in Fig.~\ref{fig:IR_ct}) and the real radiation diagrams
(Fig.~\ref{fig:IR_real}) are modified by the factor $F_\text{MSSM}$, as are most
of the loop diagrams of Fig.~\ref{fig:IR_loop}.  The only exception is diagram
(b) of Fig.~\ref{fig:IR_loop}, which involves the coupling between a neutral
Higgs and a pair of charged Goldstone bosons: the SM and MSSM couplings are
related by\footnote{Note that the $\alpha$ used here is the tree-level value,
  rather than the so-called `effective' $\alpha$ often used to account for Higgs
  mixing.}:
\begin{equation}
    A_{HGG}
  = \frac{1}{M_{H_\sm}^2} \left(\sin(\beta-\alpha)m_{h,\tree}^2 Z_{Hh}
   +\cos(\beta-\alpha) m_{H,\tree}^2 Z_{HH} \right)A_{HGG}^\sm
  \eqsep,  
\end{equation} 
-- i.e. $A_{HGG} \neq F_\mssm A_{HGG}^\sm$.  Since $m_{h,\tree}^2
\neq m_{H,\tree}^2 \neq M_{H_{\sm}}^2$, the diagrams involving the coupling
between a neutral Higgs and a pair of charged Goldstone bosons in the SM and
MSSM are not related by the same factor as the other IR divergent diagrams (or
the real correction diagrams) and the IR divergences therefore do not cancel
between the real and virtual contributions when Higgs propagator type
corrections are applied at loop level as well as at tree level.

By keeping the corrections strictly at the one-loop level this problem can of
course be avoided, resulting in an IR finite result, just as in the SM.  In this
approach, an improved Born approximation is used for the ``leading order'' form
factor $A_\impB$ -- i.e.\ the propagator-type corrections are applied to the
tree level form factors in the following manner:
\begin{equation}
    |A_{\nlo}|^2
  = |A_\impB|^2 + 2 \re{\left[A_\tree^*\Delta A\right]}
    + \delta_\text{real} |A_\tree|^2
  \eqsep, 
\end{equation} 
where $A_\tree$ is the tree level MSSM formfactor without propagator factors,
$\Delta A$ is the correction to the form factor $A$ arising from the virtual
MSSM corrections ($\matr{Z}$-factors are not applied to the loop diagrams), and
$\delta_\text{real} |A_\tree|^2$ is the correction to the form factor
resulting from the real radiation\footnote{Throughout, a capital $\Delta$
  implies an absolute correction and small $\delta$ a relative correction.}.
While this approach does avoid the problem with IR divergences, it has a
drawback because it misses the potentially large corrections arising from Higgs
mixing at the loop level (and only gives a correction for the $\cp$-even Higgs
boson decays). Several alternative approaches have been investigated, to allow
the Higgs propagator type corrections to be included at loop level as well as at
leading order while preserving an IR-finite result.
\begin{itemize}
 \item \textbf{Option 1:} Strictly speaking, the IR divergences are a
   higher-order effect -- they occur only because we are mixing orders by
   applying the $\matr{Z}$-factors at the one-loop level. The IR divergent terms can
   therefore be calculated analytically and subtracted ``by hand'':
   \begin{equation}\label{eqn:opt1} 
       |A_\impNLO|^2
     = |A_\impB|^2 + 2\re{\left[A_\impB^*\Delta A_\imp\right]}
      +\delta_\text{real} |A_\impB|^2
      -\delta_\text{sub} |A_\impB|^2, 
   \end{equation}
   where $\Delta A_\imp$ is the contribution to the form factor from vertex
   corrections to the decay of the Higgs boson (with $\matr{Z}$-factors applied
   to the loop diagrams), and $\delta_{\text{sub}}|A_\impB|^2$ is the
   analytically-calculated subtraction term used to ensure that the squared form
   factor is IR-finite.
\item \textbf{Option 2:} A second approach is to treat the `problematic' loop
  diagram (and the corresponding part of the real radiation) strictly at
  one-loop level -- no $\matr{Z}$-factors are applied to this part of the
  correction -- whilst applying Higgs propagator corrections to all other higher
  order diagrams.  In this case, 
  \begin{align}\label{eqn:opt2}
      |A_\impNLO|^2
    &=  |A_\impB|^2 + 2\re\left[A_\impB^*\Delta A'_\imp\right]
      + \delta_\text{real} |A_\impB|^2
    \nonumber\\
    &\phantom{{}={}}
      + 2\re\left[A_\tree^*\Delta A^\text{goldstone}\right]
      + \delta_\text{real}^\text{goldstone}|A_\tree|^2
    \eqsep,
  \end{align}
  where $\Delta A^{\text{goldstone}}$ is the contribution to the formfactor
  from the virtual correction shown in Fig.~\ref{fig:IR_loop}(b) \footnote{Note
    that this only involves the $\cp$-eigenstate Higgs boson $H$, as we are treating
    this particular diagram strictly at one-loop level, with no $\matr{Z}$-factors
    applied.}, $\delta_\text{real}^\text{goldstone}$ is the corresponding
  part of the real radiation and $\Delta A'_\imp$ is the contribution from all
  virtual corrections other than the loop diagram containing the $HG^+G^-$
  coupling, with $\matr{Z}$-factors applied.
\item \textbf{Option 3:} The origin of the remaining IR divergences is a
  mismatch between the $HGG$ coupling and the Higgs mass eigenstates.  By
  applying $\matr{Z}$-factors, the latter are determined at the one-loop level
  while the $HGG$ coupling is taken at tree-level. In a complete two-loop
  calculation this mismatch would be cured by one-loop corrections to the $HGG$
  vertex.  We therefore work with an \emph{effective} $HGG$ coupling, whose
  value is completely fixed by the requirement that IR divergences cancel at the
  one-loop level:
  \begin{align} 
       A_{HGG}^\eff
    &= -\frac{i e M_H^2}{2 M_W \sin\theta_W}
        \left(\matr{Z}_{Hh}\sin(\beta-\alpha) + \matr{Z}_{HH} \cos(\beta-\alpha)\right)
    \nonumber\\
    &= F_\mssm A_{H_\sm GG}
    \eqsep.\label{eqn:AHGGeff}
  \end{align}
  With this effective coupling, the NLO $HVV$ form factor is
  \begin{equation}
    \label{eqn:opt3} 
      |A_\impNLO|^2
    = |A_\impB|^2 + 2 \re\left[A_\impB^*\Delta A_\imp\right]
     + \delta_\text{real}|A_\impB|^2
    \eqsep,  
  \end{equation}
  where $\Delta A_\imp$ denotes the one-loop vertex corrections with
  $\matr{Z}$-factors applied and using the effective $HGG$ coupling from
  \eqref{eqn:AHGGeff}.
\end{itemize}

\begin{figure}
  \centering\includegraphics{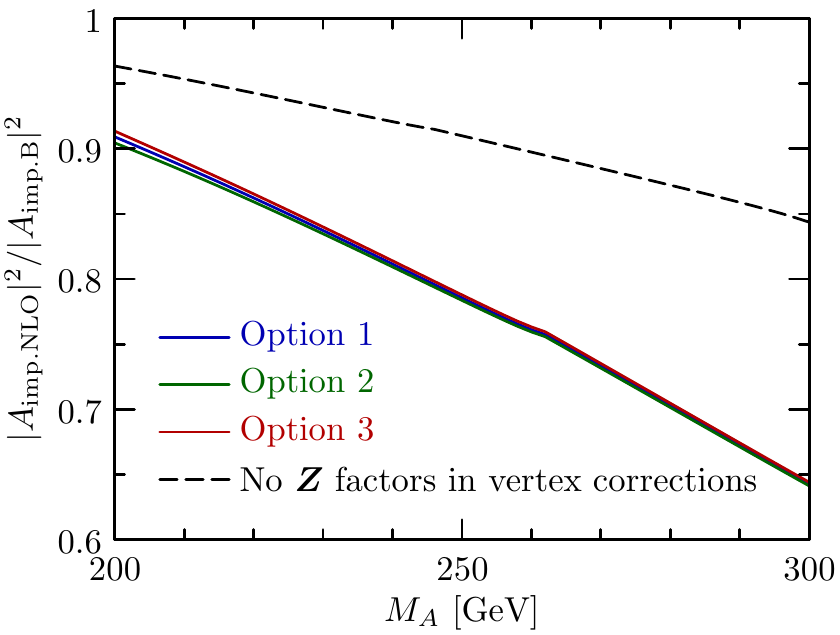}
  \caption{Numerical comparison of Options 1, 2 and 3 for dealing with IR
    divergent diagrams. Shown are the corrections to the form factor A for the
    $WW$ final state as a function of $M_A$.}
  \label{fig:comp_prop}
\end{figure}

The above options are compared in Figure~\ref{fig:comp_prop}, which shows the
relative correction to the formfactor $|A|^{2}$ (as given by Equations
\eqref{eqn:opt1},\eqref{eqn:opt2},\eqref{eqn:opt3}) as a function of the mass of
the $\cp$-odd Higgs boson, $M_{A}$.  The other parameters used are those given
in Eq.~\eqref{eq:benchmark}.\footnote{Plots varied over different parameters,
  and around a different point in the MSSM parameter space, show very similar
  features.} As can be seen, the effect of applying propagator corrections at
loop level as well as at leading order is significant. To obtain reliable
results for the effective couplings the propagator corrections should therefore
also be included at loop level. The curves for Options 1-3, where different
methods are used to eliminate the IR divergences, are very similar to one
another -- i.e. there is little practical difference between the three
approaches.

In the following, we choose to use Option 3 -- we make use of an altered
coupling between the Higgs boson and a pair of charged Goldstone bosons and add
the real radiation given by Eq.~\eqref{eq:realrad} (with the appropriate MSSM
coupling constants) to render the process IR-finite.  By using this option, we
can calculate the ratio of NLO form factors
\begin{equation}
    \frac{|A_\impNLO|^2}{|A^\sm_\nlo|^2}
  = \frac{|A_\impB|^2}{|A^\sm_\tree|^2}\cdot
    \frac{1+\frac{2\re{\left[A_\impB^*\left( \Delta A_\virt
                                            +\Delta A_\gamma\right)\right]}}
                 {|A_\impB|^2}+\delta_\real}
         {1+\frac{2\re{\left[\left(A^{\sm}_\tree\right)^{*}
                  \left(\Delta A^\sm_\virt+\Delta A^\sm_\gamma\right)\right]}}
                 {|A^\sm_\tree|^2}+\delta_\text{real}}
\end{equation}
where $\Delta A_\gamma$ is the correction to the form factor from the virtual
diagrams involving a photon (shown in Figures~\ref{fig:IR_loop},\ref{fig:IR_ct}) and $\Delta A_\virt$ symbolizes all other
virtual diagrams, so that
\begin{equation}
  \Delta A_\gamma + \Delta A_\virt = \Delta A_\imp
  \eqsep.
\end{equation}
Using the relations between the SM and MSSM form factors, we can see that, when
the effective $HGG$ coupling is used:
\begin{align}
    A_\impB
  &=F_\mssm A^\sm_\tree
  \nonumber\\
    \Delta A_\gamma
  &=F_\mssm\Delta A^\sm_\gamma
  \nonumber\\
    \frac{2\re{\left[A_\impB^*\Delta A_\gamma\right]}}{|A_\impB|^2}
  &=\frac{2\re{\left[\left(A^{\sm}_\tree\right)^{*}\Delta A^\sm_\gamma\right]}}
         {|A^\sm_\tree|^2}
   \equiv \delta^\gamma
\end{align}
Expanding, the ratio becomes
\begin{align}
    \frac{|A_\impNLO|^2}{|A^\sm_\nlo|^2}
  &=\frac{|A_\impB|^2}{|A^\sm_\tree|^2}
    \left(\frac{1 + \delta^\gamma + \delta_\virt + \delta_\real}
         {1 + \delta^\gamma + \delta^\SM_\virt + \delta_\real} \right)
  \nonumber\\
  &=\frac{|A_\impB|^2}{|A^\sm_\tree|^2}
    \left(1 + \delta_\virt - \delta^\sm_\virt\right)
  \eqsep,\label{eq:ratio}
\end{align}
where
\begin{equation}
   \delta_\virt
  =\frac{2\re{\left[A_\impB^*\Delta A_\virt\right]}}{|A_\impB|^2}
  \eqsep,\eqsep
   \delta_\virt^\sm
  =\frac{2\re{\left[\left(A^{\sm}_\tree\right)^{*}\Delta A_\virt^\sm\right]}}{|A_\tree^\sm|^2}
  \eqsep.
\end{equation}
Note that contributions from IR divergent diagrams cancel in the ratio of form
factors. This cancellation happens only when we use Option 3, i.e.\ when we use
an effective $HGG$ coupling to eliminate two-loop IR divergent terms\footnote{Of
  course the mass of the SM Higgs boson $M_{H_{\sm}}$ must be set equal to 
  the heavy MSSM Higgs boson mass $M_H$.}.

To include the contribution of the formfactor $B$ we simply note that the ratio
of polarisation sums of the $AB$ interference term and the $A^2$ term
(cf. Eq.~\eqref{eq:formfac}) is
\begin{equation}
   r_{WW} = \frac{\beta_0^2(1+\beta_0^2)}{3\beta_0^4-2\beta_0^2+3}m_H^2
\end{equation}
with $\beta_0$ from Eq.\eqref{eq:beta0}. There is no interference between the
formfactor $C$ and the other two form factors. The ratio of partial widths
may thus be written as
\begin{equation}\label{eq:rhoWW}
   \rho_{WW}
  =\frac{\Gamma}{\Gamma^\sm}
  =\frac{|A_\impB|^2}{|A^\sm_\tree|^2}
   \left[1 + \delta_\virt - \delta^\sm_\virt
           + r_{WW}(\delta'_\virt - \delta'^\sm_\virt)\right]
\end{equation}
with
\begin{equation}
   \delta'_\virt
  =\frac{2\re{\left[A_\impB^*B\right]}}{|A_\impB|^2}
  \eqsep,\eqsep
  \delta'^\sm_\virt
  =\frac{2\re{\left[\left(A^{\sm}_\tree\right)^{*} B^\sm\right]}}{|A_\tree^\sm|^2}
  \eqsep.
\end{equation}
An analogous relation holds for the ratio of $H\to ZZ$ partial widths.  We have
of course checked that the full ratio, including all contributions from photon
diagrams, agrees well with the ratio as given in Eq.~\eqref{eq:ratio}, and that
all divergences (IR and UV) cancel in the full expression for the partial width.

For off-shell decays the ratio $\rho_{Vff'}$ ($V=W,Z$) can be calculated with
only a few modifications to the expressions above. First of all, the form
factors $A$ and $B$ have to be calculated with one external gauge boson mass
replaced by $M_{ff'}$ (the invariant mass of the fermion pair). Furthermore, the
ratio $r_{Vff'}$ of polarisation sums has a more complicated form:
\begin{equation}
   r_{Vff'}
  =\frac{(M_H^2 - (M_V + M_{ff'})^2)(M_H^2 - (M_V - M_{ff'})^2)
         (M_H^2 - M_V^2 - M_{ff'}^2)}
        {2(M_H^4 + M_V^4 + 10 M_V^2 M_{ff'}^2 + M_{ff'}^4
                 - 2 M_H^2 (M_V^2 + M_{ff'}^2))}
  \eqsep.
\end{equation}
With these modifications, the expression Eq.~\eqref{eq:rhoWW} can also be used
for off-shell decays. In particular, the contributions from IR divergent
diagrams still cancel in the ratio of (differential) partial widths, as long as
Option 3 is used for the combination of propagator-type corrections and vertex
corrections.

%%%%%%%%%%%%%%%%%%%%%%%%%%%%%%%%%%%%%%%%%%%%%%%%%%%%%%%%%%%%%%%%%%%%%%%%%%%%%%%%
\section{The Parameter Scan}
\label{sec:parscan}
%%%%%%%%%%%%%%%%%%%%%%%%%%%%%%%%%%%%%%%%%%%%%%%%%%%%%%%%%%%%%%%%%%%%%%%%%%%%%%%%

As previously stated, important higher-order corrections to $H\to WW,ZZ$ decays
come from self-energy corrections of the initial-state Higgs boson (i.e.\ from
the $\matr{Z}$-factor contributions). These self-energy corrections also modify
the Higgs boson mass and are dominated by loop diagrams involving the top Yukawa
coupling, i.e.\ loops of top quarks and squarks. Beyond leading order, the $H\to
WW,ZZ$ decay rates therefore depend mainly on those MSSM parameters that enter
the Higgs-stop-stop couplings and the stop mass matrix. These parameters are
$\tan\beta$, the Higgsino mass parameter $\mu$, $A_t$ (the stop trilinear
coupling), $m_{\tilde t_R}$ (the right-handed stop mass) and $m_{\tilde q_{L3}}$
(the soft mass of the left-handed third-generation squarks).

For small $\tan\beta$ the production of MSSM Higgs bosons at the LHC proceeds
mainly through the loop-induced process $gg\to H$. In the SM this process is
mediated by a quark loop. In the MSSM squarks can also appear in the
loop. Again, the relevant contributions come from the diagrams involving the top
Yukawa coupling. For large $\tan\beta$ the coupling of $H$ to down-type quarks
is enhanced by a factor $1/\cos\beta$, which can make the $b\bar b\to H$
production mode dominant \cite{Barnett:1987jw,Dicus:1988cx,Dittmaier:2003ej}. In
any case, the relevant MSSM parameters for the most important Higgs production
processes are also $\tan\beta$, $\mu$, $A_t$, $m_{\tilde t_R}$ and $m_{\tilde
  q_{L3}}$.

These parameters are constrained by a number of experimental bounds. The
stron\-gest constraint comes from the observation of a Higgs-like resonance
\cite{ATLAS:2012gk, CMS:2012gu}, which we take to be the lightest MSSM Higgs
boson. As the MSSM tree-level relations predict a light Higgs mass below the $Z$
mass, the loop corrections have to push this mass up to \unit{126}{GeV}. The
dominant contributions to the Higgs mass still come from top and stop loops, but
contributions from other sectors can also be relevant for satisfying the
experimental bounds. Hence, the observation of a light Higgs boson at
\unit{126}{GeV} constrains many MSSM parameters simultaneously in a non-trivial
way. Thus, to include the bounds correctly, we also have to consider MSSM
parameters that have no significant impact on the heavy Higgs production or
decay rates. These parameters include the gaugino masses and the soft masses of
superpartners of the light fermions. Maximising the $pp\to H\to WW,ZZ$ cross
sections in the allowed part of this high-dimensional parameter space ``by
hand'' would be both difficult and error-prone. We therefore rely on the
numerical method of \emph{adaptive parameter scans} as suggested in
\cite{Brein:2004kh}. In the following paragraphs we describe our setup and the
scanning method in detail.

To determine the largest possible $pp\to H\to WW,ZZ$ cross sections within the
MSSM we scan over the following set of independent parameters:
\begin{itemize}
\item $\tan\beta$, $\mu$ and $M_{A}$,
\item the gaugino masses $M_1$, $M_2$ and $M_3$,
\item the stop trilinear coupling $A_t$\footnote{The trilinear couplings $A_{b}$
and $A_{\tau}$ are set to $A_{t}$, and the trilinear couplings for the first and
second fermion generations are set to zero},
\item a universal soft mass $m_{\tilde l}$ for sleptons and sneutrinos,
\item a common soft mass $m_{\tilde q}$ for all squarks \emph{except} for the
  right-handed stop mass $m_{\tilde t_R}$,
\item the mass $m_{\tilde t_1}$ of the light top squark, obtained by a judicious
  choice of the soft mass $m_{\tilde t_R}$ for right-handed stops after all
  other parameters are fixed.
\end{itemize}
This scenario is a subclass of the so-called \emph{phenomenological MSSM}
\cite{Djouadi:2002ze}.  It assumes all parameters of the soft SUSY Lagrangian to
be flavour-diagonal, which is justified by the fact that new flavour structures
in the MSSM Lagrangian are strongly constrained by flavour physics and thus have
no significant effect in Higgs physics. Searches for SUSY particles at the LHC
indicate that the superpartners of light quarks must be heavier than
approximately \unit{1}{TeV}. For simplicity, we use a common mass scale
$m_{\tilde q}$ for the corresponding soft masses. The LHC bounds on
slepton and sneutrino masses are much weaker, so we use a different scale,
$m_{\tilde l}$, for the soft masses of sleptons and sneutrinos. In fact, if the
anomalous magnetic moment of the muon $(g-2)_\mu$ is used as an experimental
bound, it is crucial to have a lower mass scale for sleptons and
sneutrinos. Furthermore, the current LHC data cannot exclude a top squark that
is lighter than the top quark. Thus, we keep the parameters $A_t$ and $m_{\tilde
  t_R}$ independent, since they only enter the stop mass matrix. For
convenience, we then trade the soft mass $m_{\tilde t_R}$ of the right-handed
stop for the physical mass $m_{\tilde t_1}$ of the lightest stop.

Important constraints on this 10-dimensional parameter space come not only from
direct Higgs searches at LEP, Tevatron and LHC, but also from the anomalous
magnetic moment $a_\mu$ of the muon, the branching ratio $\BR(B\to X_s\gamma)$
and electroweak precision observables such as the $\rho$ parameter, the
effective leptonic mixing angle $\theta_l^\text{eff}$ and the $W$ mass. Thus, we
discard portions of the parameter space according to the following criteria:
\begin{itemize}
\item We discard any set of parameters that is excluded by direct Higgs searches
  at LEP, Tevatron and LHC at 95\% CL. To do this, the masses and decay widths
  of the MSSM Higgs bosons and their effective couplings to SM fermions and
  gauge bosons are calculated with \code{FeynHiggs 2.7.4} \cite{9812320,
    9812472, 0212020, 0611326, 07050746, CPC180, Hahn:2010te}. The effective
  couplings are (for the most part) implemented in the improved Born
  approximation, i.e.\ using higher-order corrections to the Higgs self-energies
  but no vertex corrections.\footnote{Details about this approximation can be
    found in \cite{0611326}.} The masses, decay widths and effective couplings
  are then passed to \code{HiggsBounds 3.8.0} \cite{Bechtle:2008jh,
    Bechtle:2011sb}, which confronts this information with 426 different search
  channels at LEP, Tevatron and LHC \cite{arXiv:1202.1416, arXiv:1204.2760,
    arXiv:1011.1931, hep-ex/0206022, arxiv:1202.1488, arXiv:1108.5064,
    arXiv:0906.1014, arXiv:1008.3564, arXiv:1202.1415, arXiv:1109.3357,
    arXiv:1005.3216, arXiv:1106.4782, hep-ex/0107032, arXiv:0911.3935,
    arXiv:1109.3615, arXiv:1001.4468, arXiv:1108.3331, hep-ex/0401022,
    arXiv:0712.0598, arxiv:1202.1997, arXiv:0912.5285, arXiv:0707.0373,
    hep-ex/0107031, arXiv:0907.1269, arXiv:1106.4885, arXiv:0806.0611,
    arXiv:0905.3381, arXiv:1003.3363, arXiv:1106.4555, hep-ex/0111010,
    arxiv:1202.1489, arXiv:1107.1268, hep-ex/0501033, arXiv:1107.4960,
    arXiv:1012.0874, arXiv:1107.5003, arxiv:1112.2577, arXiv:0809.3930,
    arXiv:1202.1414, arXiv:0906.5613, hep-ex/0602042, arXiv:0908.1811,
    arxiv:1202.1408, arXiv:0903.4800, arXiv:1001.4481, hep-ex/0404012,
    arXiv:1107.5518, arxiv:1202.3478, arXiv:0901.1887, hep-ex/0410017, CDFnotes,
    D0notes, CMSnotes, ATLASnotes, LHWGnotes}. Internally, \code{HiggsBounds}
  uses a number of Standard Model results for the Higgs sector
  \cite{hep-ph/9704448, hep-ph/0102227, hep-ph/0102241, hep-ph/0201206,
    hep-ph/0207004, hep-ph/0302135, arXiv:0811.3458, Dawson:1990zj,
    Djouadi:1991tka, hep-ph/9504378, hep-ph/0404071, hep-ph/0407249,
    arXiv:0809.1301, arXiv:0809.3667, hep-ph/0306211, arXiv:0901.2427,
    hep-ph/0307206, hep-ph/0306234, hep-ph/0406152, hep-ph/0304035,
    hep-ph/9206246, hep-ph/9905386, hep-ph/0306109, hep-ph/0403194,
    hep-ph/0612172, hep-ph/0107081, hep-ph/0107101, hep-ph/0211438,
    hep-ph/0305321, arXiv:0705.2744, arXiv:0707.0381, arXiv:0710.4749,
    arXiv:1101.0593} to convert between experimental limits with
  different normalisations. Parameter sets that are excluded at 95\% CL by any
  of these searches are discarded.
\item To account for the discovery of a resonance at \unit{126}{GeV}
  \cite{ATLAS:2012gk, CMS:2012gu} we discard any parameter sets for which the
  light MSSM Higgs boson mass lies outside the interval between \unit{123}{GeV}
  and \unit{129}{GeV}.
\item As pointed out earlier, searches for supersymmetric particles at the LHC
  have already put strong constraints on the masses of squarks and gluinos.
  However, the interpretation of the individual searches in the context of a
  generic MSSM scenario is far from trivial and beyond the scope of this work.
  Thus, we simply require the gluinos and squarks of the first two generations
  to be heavier than \unit{1}{TeV}.  Note that the squark mass limits do not
  apply to light stops due to their different production mechanism and decay
  pattern. For $m_{\chi^0_1}>\unit{80}{GeV}$ there is currently no lower limit
  on $m_{\tilde t_1}$ from LHC. For top squarks and all uncoloured
  supersymmetric particles we therefore use (largely model-independent) mass
  limits from LEP and Tevatron, as detailed in the SUSY review of
  \cite{Nakamura:2010zzi}.  Specifically, we require that
  \begin{itemize}
  \item all slepton masses are larger than \unit{100}{GeV},
  \item all chargino masses must be larger than \unit{90}{GeV},
  \item the gluino and all squark masses except $m_{\tilde t_1}$ must be
    larger than \unit{1}{TeV},
  \item $m_{\tilde t_1}>\unit{100}{GeV}$.
  \end{itemize}
\item The MSSM contributions $\Delta a_\mu$ to the anomalous magnetic moment
  $a_\mu=(g-2)_\mu/2$ of the muon are compared with the discrepancy between the
  SM prediction and the experimental value. We require \cite{Nakamura:2010zzi}
  \begin{equation}
%    9.5\times10^{-10} \leq \Delta a_\mu \leq 41.5 (Patricks reference)
    2\times10^{-10} \leq \Delta a_\mu \leq 36\times10^{-10}
    \eqsep.
  \end{equation}
  The $2\sigma$ range was extended by the uncertainty of the SM prediction for
  $a_\mu$.
\item The MSSM contributions $\Delta\rho$ to the $\rho$ parameter are restricted
  to \cite{Nakamura:2010zzi}
  \begin{equation}
    -0.0007 \leq \Delta\rho \leq 0.0033
    \eqsep.
  \end{equation}
\item After including MSSM corrections, the effective leptonic mixing angle
  $\theta_l^\text{eff}$ is required to satisfy \cite{Nakamura:2010zzi}
  \begin{equation}
    0.2280 \leq \sin^2\theta_l^\text{eff} \leq 0.2352
    \eqsep.
  \end{equation}
\item With MSSM corrections included, the $W$ mass is required to satisfy
  \cite{Nakamura:2010zzi}
  \begin{equation}
    \unit{80.358}{GeV} \leq M_W \leq \unit{80.482}{GeV}
    \eqsep.
  \end{equation}
\item For the branching ratio $\BR(B\to X_s\gamma)$ we impose very conservative
  limits. Our reasoning is that, unlike the other low-energy observables,
  $\BR(B\to X_s\gamma)$ is very sensitive to small violations of the assumption
  of minimal flavour violation. By introducing new flavour structures in the
  MSSM we can therefore compensate deviations of $\BR(B\to X_s\gamma)$ from its
  experimental value while leaving all other observables essentially unaltered.
  Thus, we conservatively require
  \begin{equation}
    2.5\times 10^{-4} \leq \BR(b\to s\gamma) \leq 5.5\times 10^{-4}
    \eqsep.
  \end{equation}
  The correlation between the $pp\to H\to VV$ cross sections and $\BR(b\to
  s\gamma)$ will be discussed in more detail later on.
\item In the vicinity of two-particle thresholds fixed-order calculations are
  numerically unstable and resummation techniques are required to obtain
  reliable results. To avoid this kind of numerical instability we discard all
  parameter sets where the mass of the Higgs boson $H$ is within \unit{2}{GeV}
  of the sum of masses of two particles that it couples to directly. This is
  not a physical constraint but merely a precaution to stop the scan algorithm
  from running into regions where our calculations are not reliable.
\end{itemize}
MSSM corrections to the low-energy observables are calculated with
\code{FeynHiggs~2.7.4}. 

As mentioned above, we use the numerical method described in \cite{Brein:2004kh}
to systematically search for regions of the (experimentally allowed) parameter
space where $R_{VV}$ (as defined in Eq.~\eqref{eq:RVV}) is large. To this end,
we must define an \emph{importance function} $G(\vec x)$, which is a real-valued
function of the unknown model parameters:
\begin{equation}
  \vec x = (\tan\beta,M_{A},M_1,M_2,M_3,A_t,\mu,m_{\tilde l},m_{\tilde q,}
            m_{\tilde t_1})
  \eqsep.
\end{equation}
We then use the VEGAS algorithm \cite{VEGAS} to compute the integral of $G$ over
$\vec x$.  For the numerical integration we employed a modified version of the
\code{OmniComp-Dvegas} package \cite{DVEGAS}, which facilitates parallelised
adaptive Monte Carlo integration and was developed in the context of
\cite{Kauer:2001sp,Kauer:2002sn}.  With each call to the integrand function, the
parameters $\vec x$, the value of $R_{VV}$ and other relevant quantities are
written to a file. This data may then be used to study the allowed range of
$R_{VV}$ and its correlation with other parameters and observables. If the
importance function is chosen in a suitable way, the adaptive nature of the
VEGAS algorithm guarantees that the ``interesting'' regions of the parameter
space, i.e.\ those which exhibit relatively large values of $R_{VV}$, are
sampled with a higher density. Note that neither the function $G$ nor its
integral have any physical meaning. The sole purpose of the importance function
is to drive the adaptation of the VEGAS algorithm into those regions of
parameter space we are interested in. An obvious choice for $G(\vec x)$ would
therefore be
\begin{equation}
  G(\vec x) =
  \begin{cases} R_{VV}(\vec x), & \text{$\vec x$ satisfies constraints} \\
                0,              & \text{otherwise}
  \end{cases}
  \eqsep,
\end{equation}
where the `constraints' are those discussed earlier in this section. The
performance of the algorithm can be improved by using exponential dampening
instead of ``hard cuts'' in the importance function. Thus, we write $G(\vec x)$
as
\begin{equation}
  G(\vec x) = R_{VV}(\vec x)\prod_i y_i(O_i(\vec x))
  \eqsep,
\end{equation}
where the $O_i$ denote all the observables used to constrain the parameter
space. The functions $y_i$ are chosen to be equal to one for allowed values of
the corresponding observable $O_i$ and to drop off exponentially outside the
allowed range. In our scan, the constraints from the low-energy observables
$\Delta a_\mu$, $\Delta\rho$, $\sin^2\theta_l^\text{eff}$, $M_W$ and $\BR(b\to
s\gamma)$ are implemented in this way. The constraints from Higgs searches are
also treated in this manner. The related `observable' is the ratio $S_{95}$ of
the signal cross section divided by the observed 95\% CL limit for the most
sensitive search channel, as provided by \code{HiggsBounds}. The mass bounds on
SUSY particles, on the other hand, are implemented as hard cuts. Thus, our
importance function may be non-zero for some points that do not satisfy the
constraints discussed earlier. However, the scatter plots we show in this paper
only contain points that satisfy the constraints.

%%%%%%%%%%%%%%%%%%%%%%%%%%%%%%%%%%%%%%%%%%%%%%%%%%%%%%%%%%%%%%%%%%%%%%%%%%%%%%%%
\section{Results and discussion}
\label{sec:results}
%%%%%%%%%%%%%%%%%%%%%%%%%%%%%%%%%%%%%%%%%%%%%%%%%%%%%%%%%%%%%%%%%%%%%%%%%%%%%%%%

In the first part of this section we discuss scatterplots of
$R_{VV}^\text{imp.B}$, as defined in Eq.~\eqref{eq:RVV} and evaluated in the
improved Born approximation, against different input parameters and
observables. The density of the points in these plots has no
\textit{statistical} interpretation. It is, however, safe to say that regions
with a very low density of points can only be realised with rather finely tuned
parameters.

\begin{figure}
  \centering
  \includegraphics{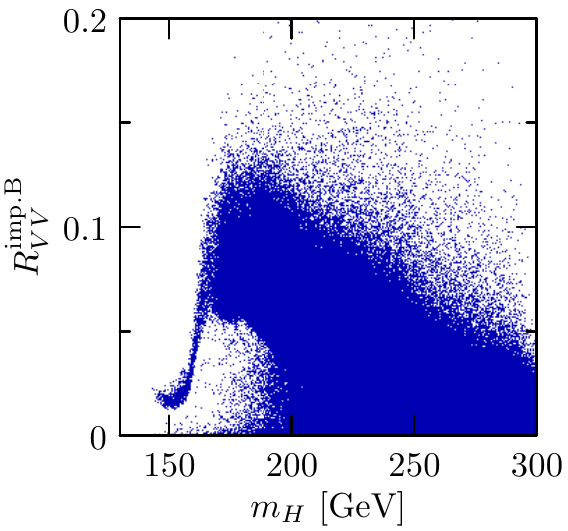}
  \caption{Scatterplot of $R_{VV}^\text{imp.B}$ (Eq.~\eqref{eq:RVV} evaluated in
    the improved Born approximation) against the heavy scalar Higgs mass $M_H$.}
  \label{fig:mh2_S0}
\end{figure}
Figure~\ref{fig:mh2_S0} shows a scatterplot of $R_{VV}^\text{imp.B}$ against the
heavy scalar Higgs mass $M_H$. We see that typical values of
$R_{VV}^\text{imp.B}$ do not exceed $0.15$. On the other hand, the experimental
results for SM $H\to WW$ and $H\to ZZ$ searches presented at the ICHEP2012
conference probe, for certain Higgs masses, values of $R_{VV}^\text{imp.B}$ as
low as $0.1$.\footnote{Note that the constraints from direct Higgs searches
  implemented in our scan (via \code{HiggsBounds}) are only based on 2011 Higgs
  data.}  In other words, the searches for heavy Higgs bosons already rule out
certain regions of the MSSM parameter space. For $M_H\lesssim\unit{160}{GeV}$
the values of $R_{VV}$ drop rapidly below $0.05$. The largest values of
$R_{VV}^\text{imp.B}$ are reached for $M_H$ between $160$ and
$\unit{200}{GeV}$. For $M_H>\unit{240}{GeV}$ the typical size of
$R_{VV}^\text{imp.B}$ drops back below $0.1$.

\begin{figure}
  \centering
  \includegraphics{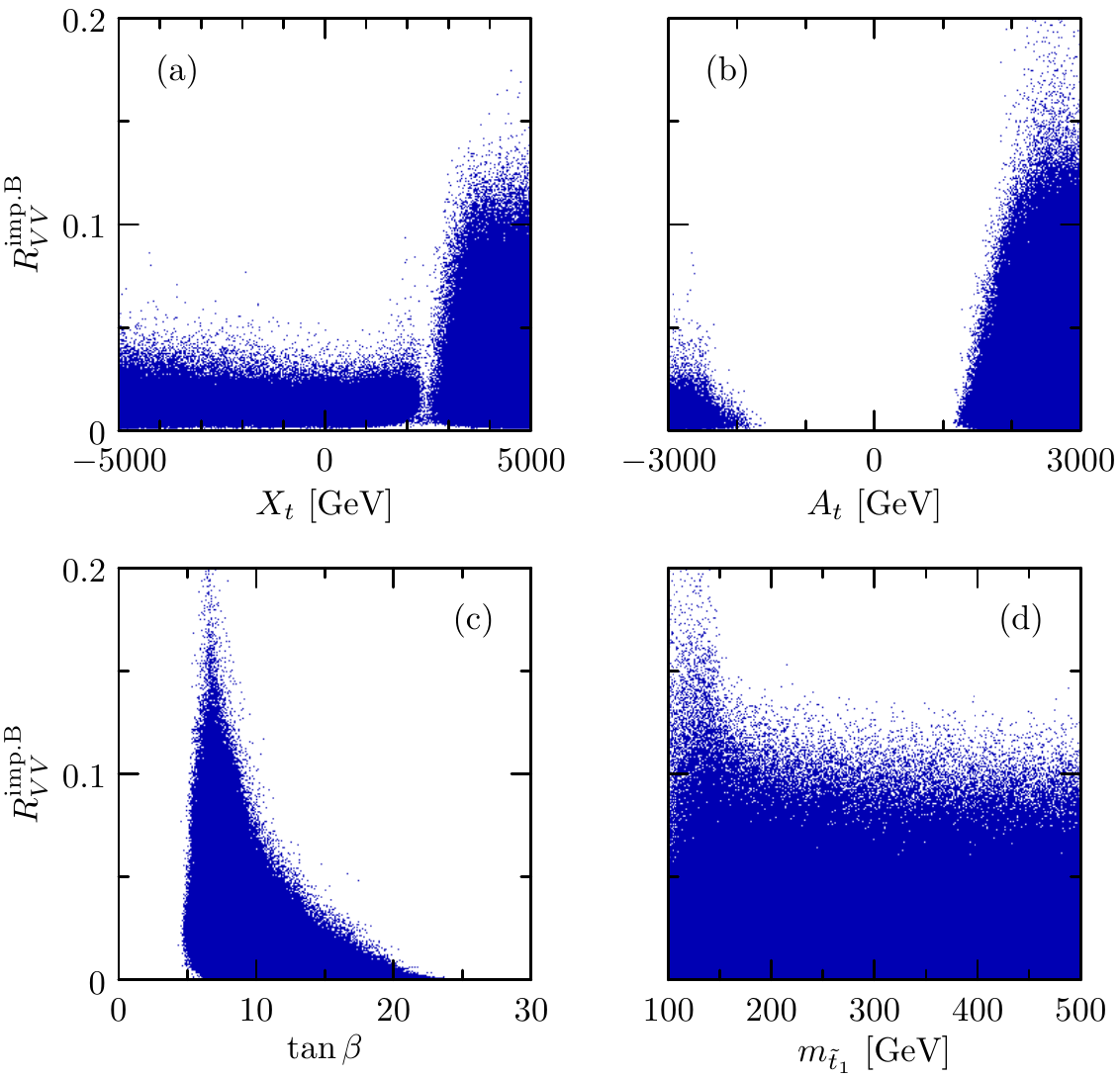}
  \caption{Scatterplots of $R_{VV}^\text{imp.B}$ (Eq.~\eqref{eq:RVV} evaluated
    in the improved Born approximation) against the parameters $A_t$,
    $\tan\beta$, $m_{\tilde t_1}$ and the quantity $X_t\equiv
    A_t-\mu\tan\beta$.}
  \label{fig:par_S0}
\end{figure}
In Figure~\ref{fig:par_S0} we show scatterplots of $R_{VV}^\text{imp.B}$ against
different input parameters. The quantity $X_t$ is defined as
\begin{equation}
  X_t = A_t - \mu\tan\beta
  \eqsep,
\end{equation}
so that $m_tX_t$ is the off-diagonal element in the stop mass-matrix.  We see
that values of $R_{VV}^\text{imp.B}$ above $0.1$ are only possible for
$A_t\gtrsim\unit{2}{TeV}$, $\tan\beta$ between approximately 5 and 10 and
$X_t\gtrsim\unit{2.5}{TeV}$. Values of $R_{VV}^\text{imp.B}$ larger than $0.15$
additionally require $m_{\tilde t_1}<\unit{150}{GeV}$.

\begin{figure}
  \centering
  \includegraphics{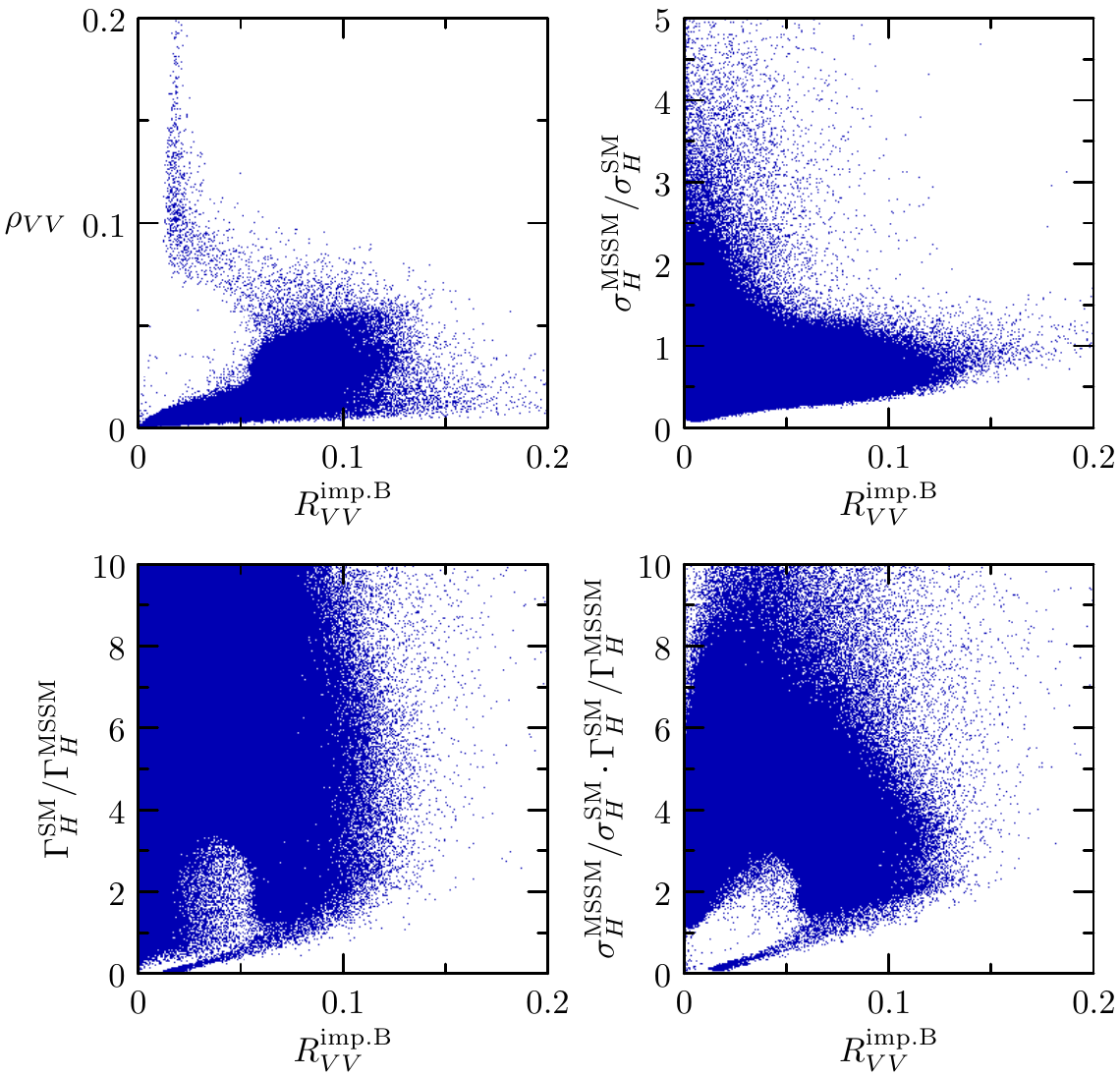}
  \caption{Scatterplots of $R_{VV}^\text{imp.B}$ (Eq.~\eqref{eq:RVV} evaluated
    in the improved Born approximation) against the three ratios $\rho_{VV}$,
    $\sigma_H^\text{MSSM}/\sigma_H^\text{SM}$ and
    $\Gamma_H^\text{SM}/\Gamma_H^\text{MSSM}$ and the product
    $\sigma_H^\text{MSSM}/\sigma_H^\text{SM}\cdot\Gamma_H^\text{SM}/
    \Gamma_H^\text{MSSM}$.}
  \label{fig:S0_rhoVV}
\end{figure}

From Eq.~\eqref{eq:RVV} we know that $R_{VV}$ is proportional to three ratios:
the ratio $\sigma_H^\text{MSSM}/\sigma_H^\text{SM}$ of production cross
sections, the ratio $\Gamma_H^\text{SM}/\Gamma_H^\text{MSSM}$ of total decay
widths (note the reversed order), and the ratio $\rho_{VV}$ of partial $H\to VV$
decay widths. The last two quantities are universal in the sense that they
appear in the signal strengths for all final states. Scatterplots of these
ratios and the product of the two universal ratios against the value of
$R_{VV}^\text{imp.B}$ are shown in Fig.~\ref{fig:S0_rhoVV}. We see that for
values of $R_{VV}^\text{imp.B}$ above $0.1$ the value of $\rho_{VV}$ usually
does not exceed $0.05$ and the remaining enhancement comes from the product of
universal factors, which, for $R_{VV}^\text{imp.B}>0.1$, typically lies between
$2$ and $4$. This is an interesting feature, since this enhancement factor would
also appear in other decays like $H\to\tau\tau$ or $H\to b\bar b$.

\begin{figure}
  \centering\includegraphics{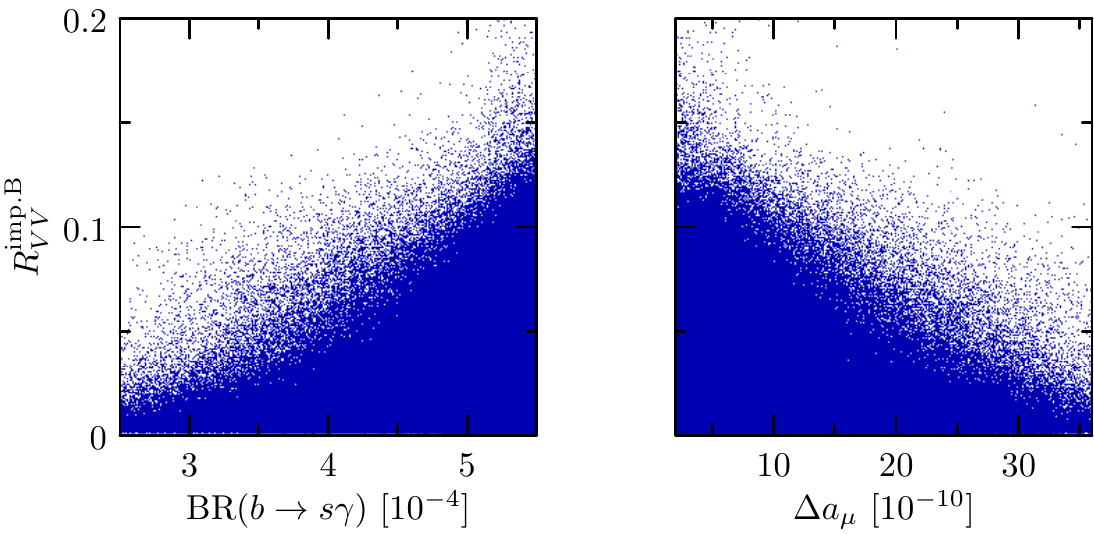}
  \caption{Scatterplots of $R_{VV}^\text{imp.B}$ (Eq.~\eqref{eq:RVV} evaluated
    in the improved Born approximation) against the low energy observables
    $\BR(b\to s\gamma)$ and $\Delta a_\mu$.}
  \label{fig:scatter-lowE_S0}
\end{figure}
Fig.~\ref{fig:scatter-lowE_S0} shows the correlation of $R_{VV}^\text{imp.B}$
with the two most constraining low-energy observables, $\BR(b\to s\gamma)$ and
$\Delta a_\mu$. We see that large values of $R_{VV}^\text{imp.B}$ typically
coincide with large values of $\BR(b\to s\gamma)$ and small values of $\Delta
a_\mu$. The branching ratio $\BR(b\to s\gamma)$ can be driven to smaller values
by the chargino-stop loop contribution. This happens if the product $\mu A_t$ is
small and negative.  On the other hand, large values of $R_{VV}^\text{imp.B}$
require $A_t$ to be large and positive and $|\mu|$ is bounded from below through
the chargino mass limit. Thus small values of $\BR(b\to s\gamma)$ do not
coincide with large values of $R_{VV}^\text{imp.B}$. A similar argument can be
given for the correlation between $R_{VV}^\text{imp.B}$ and $\Delta a_\mu$. The
value of $\Delta a_\mu$ is approximately proportional to $\tan\beta/m_{\tilde
  l}^2$. Since large values of $R_{VV}^\text{imp.B}$ require small values of
$\tan\beta$ and $m_{\tilde l}$ is bounded from below by the slepton mass limit
we do not find any points where $R_{VV}^\text{imp.B}$ and $\Delta a_\mu$ are
large simultaneously. It should be stressed that, while these arguments make the
features in Fig.~\ref{fig:scatter-lowE_S0} plausible, they do not suffice for a
quantitative explanation. The correlations in Fig.~\ref{fig:scatter-lowE_S0} are
really a result of the combination of different experimental bounds.

\begin{figure}
  \centering\includegraphics{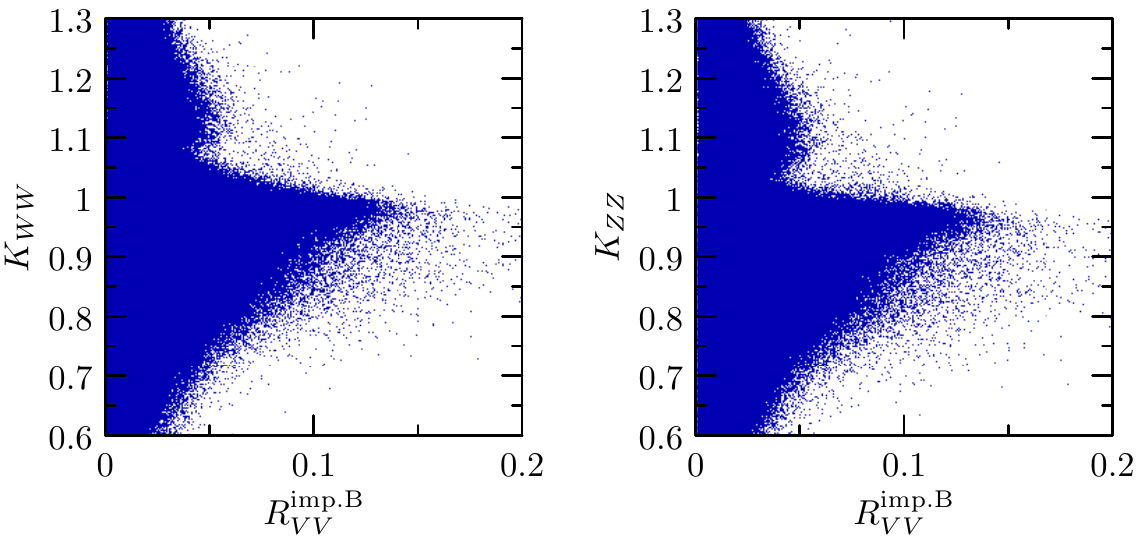}
  \caption{Scatterplots of the $K$ factors $K_{VV}$ ($V=W,Z$) due to genuine
    vertex corrections against the ratio $R_{VV}^\text{imp.B}$.}
  \label{fig:scatter-S0_KVV}
\end{figure}
To quantify the importance of the vertex corrections, we define the $K$ factors
$K_{VV}$ as
\begin{equation}
  K_{VV} = \frac{\rho_{VV}^{\mbox{\tiny{full}}}}{\rho^\impB_{VV}}
  \eqsep,
\end{equation}
where $\rho^\impB_{VV}$ denotes the ratio $\rho_{VV}$ from Eq.~\eqref{eq:RVV} in
the improved Born approximation (i.e.\ with $\matr{Z}$ factors included but
without any vertex corrections) and $\rho^{\mbox{\tiny{full}}}_{VV}$ is the
fully corrected value. Fig.~\ref{fig:scatter-S0_KVV} shows scatter plots of
$K_{WW}$ and $K_{ZZ}$ against the ratio $R_{VV}^\text{imp.B}$, i.e.\ the ratio
$R_{VV}$ calculated in the improved Born approximation. We see that, for
$R_{VV}^\text{imp.B}>0.1$ the $K$ factors due to genuine vertex corrections
typically lie between $0.8$ and $1.0$. For $R_{VV}^\text{imp.B}>0.05$ the $K$
factors can lie between $0.7$ and $1.1$. For $R_{VV}^\text{imp.B}\lesssim 0.02$
the vertex corrections are numerically as important as the results obtained in
the improved Born approximation. This means that, once the experimental
sensitivity reaches two percent of the SM signal, the vertex corrections have to
be included to derive limits on the MSSM parameters.

\begin{figure}
  \centering\includegraphics{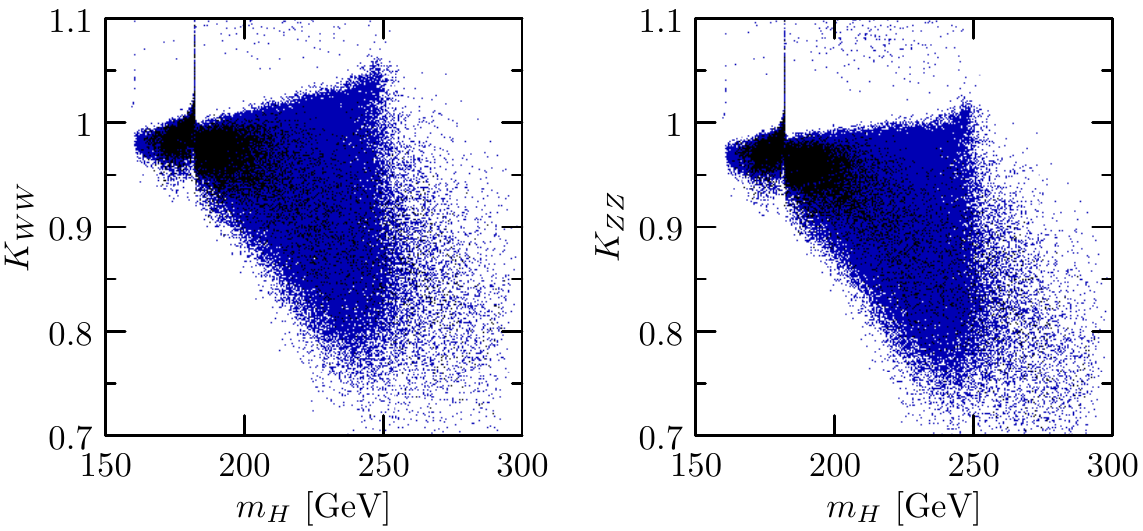}
  \caption{Scatterplots of the $K$ factors $K_{VV}$ ($V=W,Z$) due to genuine
    vertex corrections against the physical Higgs mass $M_H$. Only points with
    $R_{VV}^\text{imp.B}>0.05$ are included and points with
    $R_{VV}^\text{imp.B}>0.1$ are shown in black.}
  \label{fig:scatter-mh2_KVV}  
\end{figure}

Fig.~\ref{fig:scatter-mh2_KVV} shows scatterplots of $K_{WW}$ and $K_{ZZ}$
against the physical Higgs mass $M_H$. Only points with $R_{VV}>0.05$ are
included and points with $R_{VV}^\text{imp.B}>0.1$ are shown in black. We see
that vertex corrections with a magnitude of more than $10\%$ of the result in
the improved Born approximation appear for $M_H>\unit{200}{GeV}$ and are always
negative. The spike near $M_H=\unit{180}{GeV}$ originates from the $ZZ$
threshold.

The vertex diagrams contributing to $H\to WW$ and $H\to ZZ$ decays can be split
into three separately IR and UV finite subsets:
\begin{description}
\item[fermion/sfermion] contributions from loops containing SM fermions or their
  superpartners,
\item[chargino/neutralino] contributions from loops involving charginos or
  neutralinos and
\item[two-Higgs-doublet] contributions from loops containing only gauge,
  Goldstone and Higgs bosons.
\end{description}
To study the different contributions in more detail let us select a
representative set of parameters with a relatively large value of $R_{VV}$:
\begin{gather}
  M_{A} = \unit{260}{GeV}
  \eqsep,\eqsep
  \tan\beta = 7.5
  \eqsep,\eqsep
  A_t = \unit{1700}{GeV}
  \eqsep,\nonumber\\
  m_{\tilde l} = \unit{300}{GeV}
  \eqsep,\eqsep
  m_{\tilde q} = \unit{2000}{GeV}
  \eqsep,\eqsep
  m_{\tilde{t}_{1}} = \unit{150}{GeV}
  \eqsep,\eqsep
  m_{\tilde{g}} = \unit{1200}{GeV}
  \eqsep,\nonumber\\
  \mu = \unit{-2500}{GeV}
  \eqsep,\eqsep
  M_1 = \unit{-100}{GeV}
  \eqsep,\eqsep
  M_2= \unit{200}{GeV}
  \nonumber\\
  \Rightarrow\eqsep
  M_H = \unit{246.7}{GeV}
  \eqsep,\eqsep
  R_{WW} = 0.0854
  \eqsep,\eqsep
  R_{ZZ} = 0.0813
  \eqsep.\label{eq:benchmark}%
\end{gather}
\begin{figure}[!htb]
  \centering\includegraphics{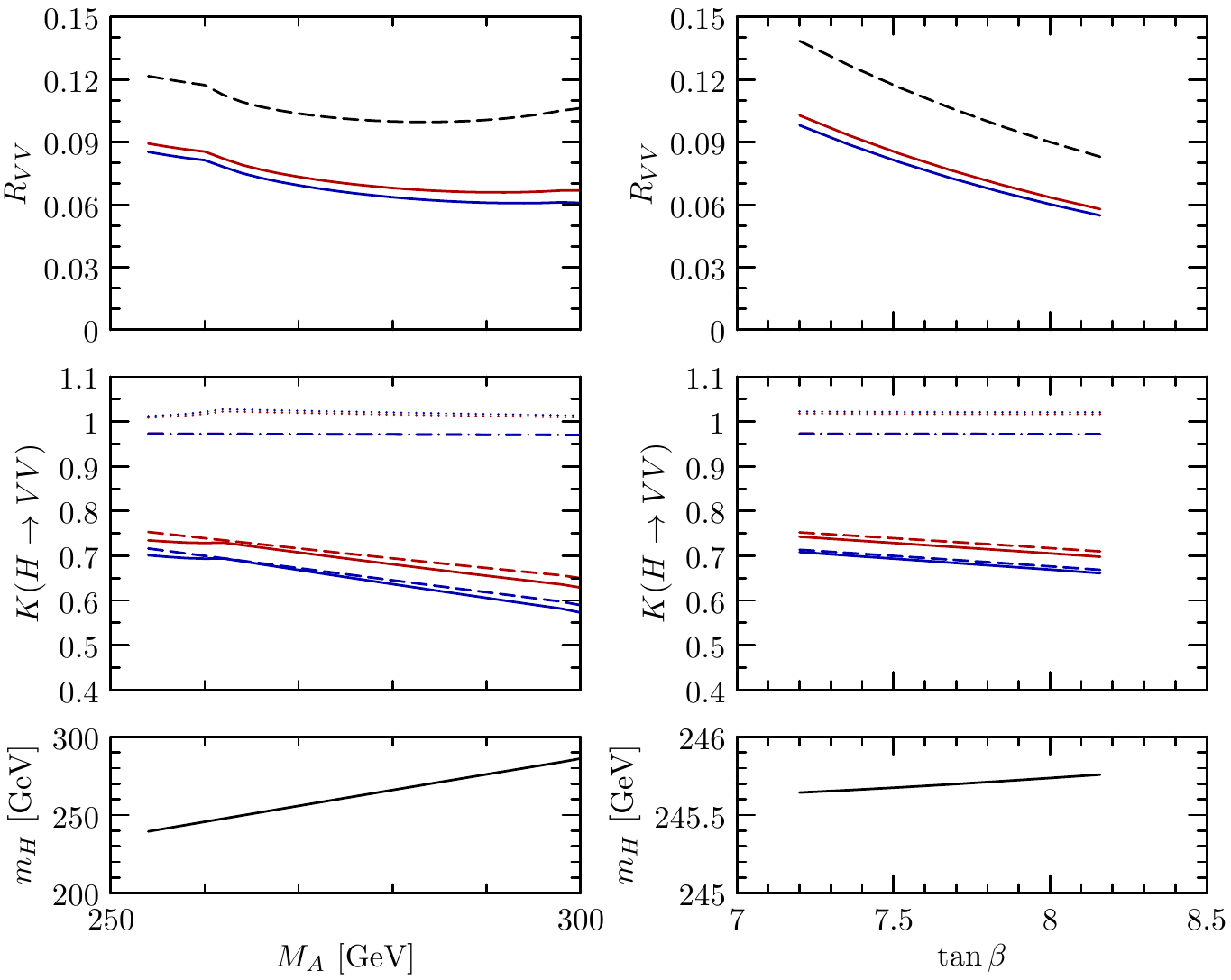}%
  \caption{Genuine vertex corrections to $R_{VV}$ ($V=W,Z$) as
    functions of $M_{A}$ (left column) and $\tan\beta$ (right column). All other
    parameters are fixed according to Eq.~\eqref{eq:benchmark}.  The plots in the
    upper row show the ratios $R_{WW}$ (red lines) and $R_{ZZ}$ (blue lines)
    with vertex corrections included. The dashed black line shows the same
    ratios in the improved Born approximation. ($R_{WW}$ and $R_{ZZ}$ are equal
    in this approximation.) The plots in the middle row show the $K$ factors due
    to fermion/sfermion vertex diagrams (dashed lines), chargino/neutralino
    diagrams (dashdotted lines) and two-Higgs-doublet model diagrams (dotted
    lines). The solid lines combine all three types of corrections. Again, red
    and blue lines represent the $WW$ and $ZZ$ final states, respectively. The
    dependence of the mass $M_H$ on $M_{A}$ and $\tan \beta$ is shown in the
    plots at the bottom. The lines are interrupted if the corresponding
    parameter point violates the constraints discussed in
    Sec.~\ref{sec:parscan}.
  }%
  \label{fig:plot-mAtb}%
\end{figure}%
The vertex corrections to $R_{WW}$ and $R_{ZZ}$ are shown in
Fig.~\ref{fig:plot-mAtb} as functions of $M_{A}$ and $\tan\beta$, with all other
parameters fixed according to \eqref{eq:benchmark}. We see that $R_{WW}$ and
$R_{ZZ}$ are different after the inclusion of vertex corrections, but the $K$
factors only differ by a few percent. The dominant vertex contributions come
from the fermion/sfermion diagrams, in particular from the diagrams involving
top quarks and stops. For the chosen scenario these contributions to the $K$
factors lie between $-30\%$ and $-40\%$. The vertex corrections from the
two-Higgs-doublet and chargino/neutralino sector typically only amount to a few
percent. For $M_{A} \approx \unit{260}{GeV}$ the $H\to hh$ threshold is crossed,
which leads to the characteristic kinks in the graph. All three types of vertex
corrections stem almost entirely from corrections to the form factor $A$ (see
Eq.~\eqref{eq:formfac}). Contributions from the loop-induced form factor $B$ to
$R_{VV}$ are of the order of $10^{-4}$ and negligible for all practical
purposes.

\begin{figure}
  \centering\includegraphics{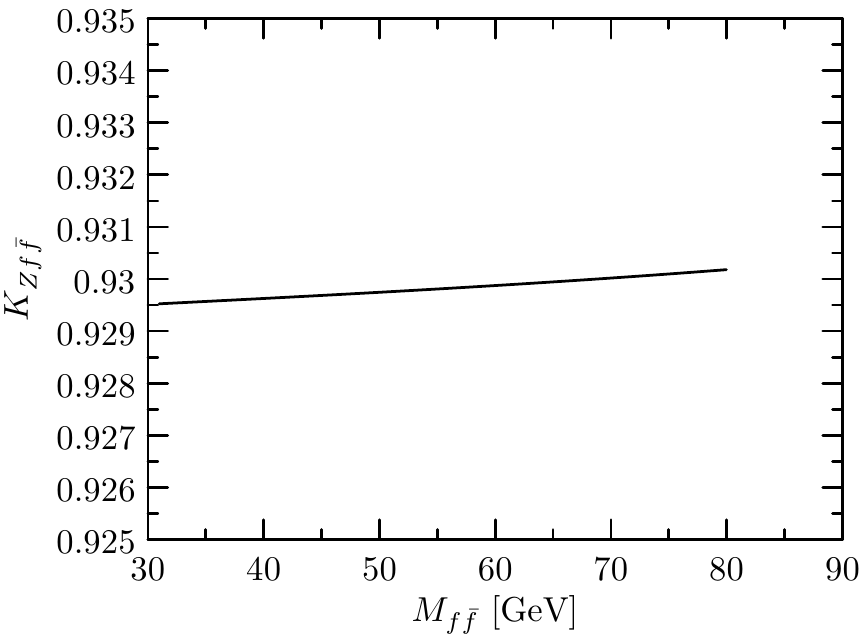}
  \caption{The off-shell $K$ factor $K_{Zf\bar f}$ as a function of the
    invariant mass $M_{f\bar f}$ of the fermion anti-fermion pair $f\bar f$ for
    the parameters of Eq.~\eqref{eq:offshell-benchmark}. The corresponding Higgs
    mass $M_H$ is \unit{170.4}{GeV}, which is below the $ZZ$ threshold.}
  \label{fig:plot-mff}
\end{figure}
Let us now examine the validity of the effective coupling approximation
in the case of off-shell decays. As explained in Sec.~\ref{sec:effc}, we can
define the off-shell $K$ factors $K_{Vff'}(M_{ff'})$ by
\begin{equation}
  K_{Vff'}(M_{ff'}) = \frac{\rho_{Vff'}(M_{ff'})}
                           {\rho_{Vff'}^\impB(M_{ff'})}
  \eqsep,
\end{equation}
where $V=W,Z$, $ff'$ may be any SM fermion pair into which $V$ can decay and
$M_{ff'}$ is the invariant mass of the $ff'$ pair. If $K_{Vff'}$ is independent
of $M_{ff'}$ (to a good approximation) the integrated MSSM cross sections for
$pp\to H\to Vff'$ can be evaluated by scaling the corresponding SM cross section
with $R_{Vff'}$ (cf.\ Eq.~\ref{eq:effc_approx}) evaluated at some fixed value of
$M_{ff'}$. Such an approximation could be useful for $M_H$ below $2M_Z$. To
study the quality of the approximation we choose a scenario where $M_H$ is below
the $ZZ$ threshold:
\begin{gather}
  M_{A} = \unit{170}{GeV}
  \eqsep,\eqsep
  \tan\beta = 8.5
  \eqsep,\eqsep
  A_t = \unit{2500}{GeV}
  \eqsep,\nonumber\\
  m_{\tilde l} = \unit{250}{GeV}
  \eqsep,\eqsep
  m_{\tilde q} = \unit{2200}{GeV}
  \eqsep,\eqsep
  m_{\tilde{t}_{1}} = \unit{250}{GeV}
  \eqsep,\eqsep
  m_{\tilde{g}} = \unit{1000}{GeV}
  \eqsep,\nonumber\\
  \mu = \unit{-900}{GeV}
  \eqsep,\eqsep
  M_1 = \unit{-200}{GeV}
  \eqsep,\eqsep
  M_2= \unit{500}{GeV}
  \nonumber\\
  \Rightarrow\eqsep
  M_H=\unit{172.2}{GeV}
  \eqsep,\eqsep
  R_{WW} = 0.102
  \eqsep,\eqsep
  R_{ZZ} = 0.100
  \eqsep.\label{eq:offshell-benchmark}%
\end{gather}
Fig.~\ref{fig:plot-mff} shows $K_{Zf\bar f}$ as a function of $M_{f\bar f}$. We
see that $K_{Zf\bar f}$ is constant within one per mille over a large range of
$M_{ff'}$. Scaling integrated SM cross sections for $pp\to H\to Vff'$ with
$R_{Vff'}$ using the approximation of Eq.~\eqref{eq:effc_approx} therefore gives
the correct MSSM cross sections with a relative accuracy of approximately one
per mille.

\begin{figure}
  \centering\includegraphics{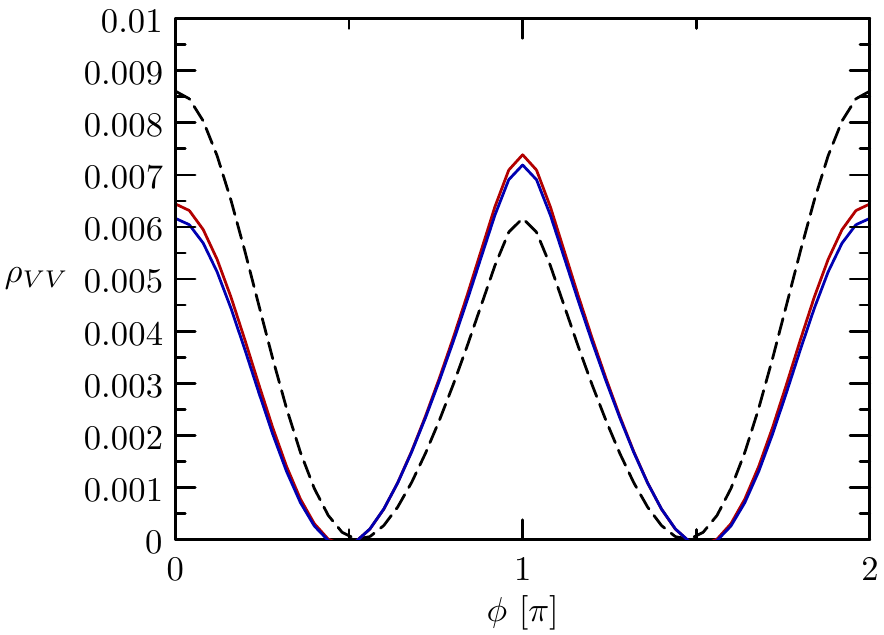}
  \caption{Dependence of the ratio $\rho_{VV}$ of partial widths on the common
    complex phase $\phi$ of the parameters $M_1$, $M_2$, $M_3$, $\mu$ and $A_t$.
    Absolute values and relative signs are chosen in such a way that we
    re-obtain the parameters of Eq.~\eqref{eq:benchmark} for $\phi=0$. The
    solid lines show the ratios $\rho_{WW}$ (red) and $\rho_{ZZ}$ (blue) with
    vertex corrections included. The dashed black line shows the same ratios in
    the improved Born approximation.}
  \label{fig:cmssm}
\end{figure}
Finally, we would like to make a few remarks regarding the complex MSSM. As
mentioned earlier, introducing $\cp$-violating phases in the MSSM Lagrangian
leads to mixing between the $\cp$-even and the $\cp$-odd neutral Higgs bosons.
We have seen that there is no interference between the form factor $C$ and the
(tree-level) form factor $A$. Thus, the largest values of $\rho_{VV}$ are
obtained if the decaying mass eigenstate has no $\cp$-odd component, i.e.\ in
the case of vanishing complex phases. This effect can be seen in
Figure~\ref{fig:cmssm}, where we take the parameter values of
Eq.~\eqref{eq:benchmark} and introduce a common complex phase $\phi$ for $M_1$,
$M_2$, $M_3$, $\mu$ and $A_t$. Absolute values and relative signs are chosen in
such a way that we re-obtain the parameters Eq.~\eqref{eq:benchmark} for
$\phi=0$.  The plot shows the dependence of the ratio $\rho_{VV}$ of partial
widths ($V=W,Z$) as a function of $\phi$. The dependence on $\phi$ can be much
weaker for different choices of MSSM parameters, which only admit smaller mixing
between the $\cp$-even and $\cp$-odd states. We have run the full parameter
scans with the complex phase $\phi$ as an additional variable, but the largest
values of $\rho_{VV}$ were achieved for $\phi=0$.

%%%%%%%%%%%%%%%%%%%%%%%%%%%%%%%%%%%%%%%%%%%%%%%%%%%%%%%%%%%%%%%%%%%%%%%%%%%%%%%%
\section{Conclusion}
\label{sec:conclusions}
%%%%%%%%%%%%%%%%%%%%%%%%%%%%%%%%%%%%%%%%%%%%%%%%%%%%%%%%%%%%%%%%%%%%%%%%%%%%%%%%

We have calculated the partial decay widths for the heavy scalar MSSM Higgs
boson $H$ decaying into $WW$ and $ZZ$ final states at one-loop order and
confirmed the available results of \cite{Hollik:2011xd}, which included the full
one-loop corrections to the $H\rightarrow ZZ$ decay width and the (s)fermion
corrections to the $H\rightarrow WW$ decay mode. We have extended the
calculation to include the full 1-loop corrections in the $H\rightarrow WW$
channel.  To improve the precision of the one-loop result we proposed a method
for combining Higgs propagator-type corrections (as defined in
\cite{0611326,07105320} and calculated by \code{FeynHiggs}) with the genuine
full one-loop vertex corrections for both $H \rightarrow ZZ$ and $H \rightarrow
WW$.  We addressed the issue of infrared divergences appearing in the $H\to WW$
process and ensured that our method leads to an IR finite result. In particular,
no IR divergent diagrams need to be evaluated in the computation of the MSSM/SM
ratio $\rho_{VV}$ ($V=W,Z$) of partial $H\to VV$ decay widths if we use a
modified coupling between the Higgs boson and a pair of charged Goldstone
bosons. The same method allows us to calculate the MSSM/SM ratios
$\rho_{Vff'}(M_{ff'})$ of differential partial widths for single off-shell
decays $H\to VV^*\to Vff'$, where $f$ and $f'$ are two massless SM fermions and
$M_{ff'}$ is their invariant mass. We find that $\rho_{Vff'}$ is independent of
$M_{ff'}$ with a relative accuracy of approximately two per mille. Partial
widths for single off-shell decay in the MSSM can therefore be safely estimated
by scaling the corresponding (off-shell) SM partial widths.

The possible size of the MSSM/SM ratios $R_{VV}$ of Higgs production cross
sections times branching ratios have been studied in an adaptive parameter scan.
Experimental constraints from several low-energy observables and direct Higgs
searches at LEP, Tevatron and LHC were included in our scan (with the help of
\code{FeynHiggs} and \code{HiggsBounds}). No assumptions about the SUSY breaking
mechanism were made. We find that $R_{VV}$ ratios of up to $0.2$ can still be
compatible with experimental constraints from direct SUSY searches and
low-energy observables for $M_H\gtrsim\unit{160}{GeV}$. These parameter space
regions are currently or will soon be probed by the direct Higgs searches at the
LHC. The one-loop vertex contributions to the decay processes typically lead to
corrections between $-30\%$ and $+10\%$ for MSSM parameters where $R_{VV}$ is
larger than $0.05$. For $R_{VV}\lesssim 0.02$ the vertex corrections can be
numerically as important as the tree-level results and Higgs self-energy
corrections and therefore have to be considered when setting limits on the MSSM
parameter space.

The source code for our calculations is available on request from M.W.

\section*{Acknowledgemnts}

We are very grateful to Georg Weiglein for helpful discussions, and for a
thorough reading of an early draft of this paper.  P.G.\ is supported by DFG
SFB/TR9.

\bibliography{references}

\end{document}